\documentclass[aps,pre,twocolumn,longbibliography,amsmath,amsmath,amssymb]{revtex4-1}

\usepackage[utf8]{inputenc}
\usepackage[english]{babel}

\usepackage{bm}

\usepackage{graphicx}

\usepackage{amsmath}
\usepackage{amsfonts}
\usepackage{amssymb}

\usepackage{algcompatible}
\usepackage{newfloat}
\DeclareFloatingEnvironment[
    fileext=loa,
    listname=List of Algorithms,
    name=ALGORITHM,
    placement=tbhp,
]{algorithm}

\usepackage[hyperindex,colorlinks,hyperfootnotes = false,citecolor = blue]{hyperref}
\usepackage[usenames,dvipsnames,svgnames,table]{xcolor}

\newcommand{\argmin}{\arg\!\min}
\newcommand{\argmax}{\arg\!\max}

\newcommand{\R}{\mathbb{R}}      % for Real numbers
\usepackage{booktabs}

\usepackage{natbib}

\makeatletter
\newcounter{savesection}
\newcounter{apdxsection}
\renewcommand\appendix{\par
  \setcounter{savesection}{\value{section}}%
  \setcounter{section}{\value{apdxsection}}%
  \setcounter{subsection}{0}%
  \gdef\thesection{\@Alph\c@section}}
\newcommand\unappendix{\par
  \setcounter{apdxsection}{\value{section}}%
  \setcounter{section}{\value{savesection}}%
  \setcounter{subsection}{0}%
  \gdef\thesection{\@arabic\c@section}}
\makeatother

\begin{document}

\title{Collective search with finite perception: transient dynamics and search efficiency} 

\begin{abstract}
Motile organisms often use finite spatial perception of their surroundings to navigate and search their habitats. Yet standard models of search are usually based on purely local sensory information. To model how a finite perceptual horizon affects ecological search,
we propose a framework for optimal navigation that combines concepts from random walks and optimal control theory. 
We show that, while local strategies are optimal on asymptotically long and short search times, finite perception yields faster convergence and increased search efficiency over transient time scales relevant in biological systems. 
The benefit of the finite horizon can be maintained by the searchers tuning their response sensitivity to the length scale of the stimulant in the environment, and is enhanced when the agents interact as a result of increased consensus within subpopulations. 
Our framework sheds light on the role of spatial perception and transients in search movement and collective sensing of the environment. 
\end{abstract}

\author{Adam Gosztolai}
\email{adam.gosztolai@unige.ch}
\affiliation{Department of Mathematics, Imperial College London, United Kingdom}

\author{Jose A. Carrillo}
\affiliation{Department of Mathematics, Imperial College London, United Kingdom}

\author{Mauricio Barahona}
\email{m.barahona@imperial.ac.uk}
\affiliation{Department of Mathematics, Imperial College London, United Kingdom}

\date{\today}

\maketitle

\section*{Introduction}

\begin{figure}
	\includegraphics[width = 0.8 \columnwidth]{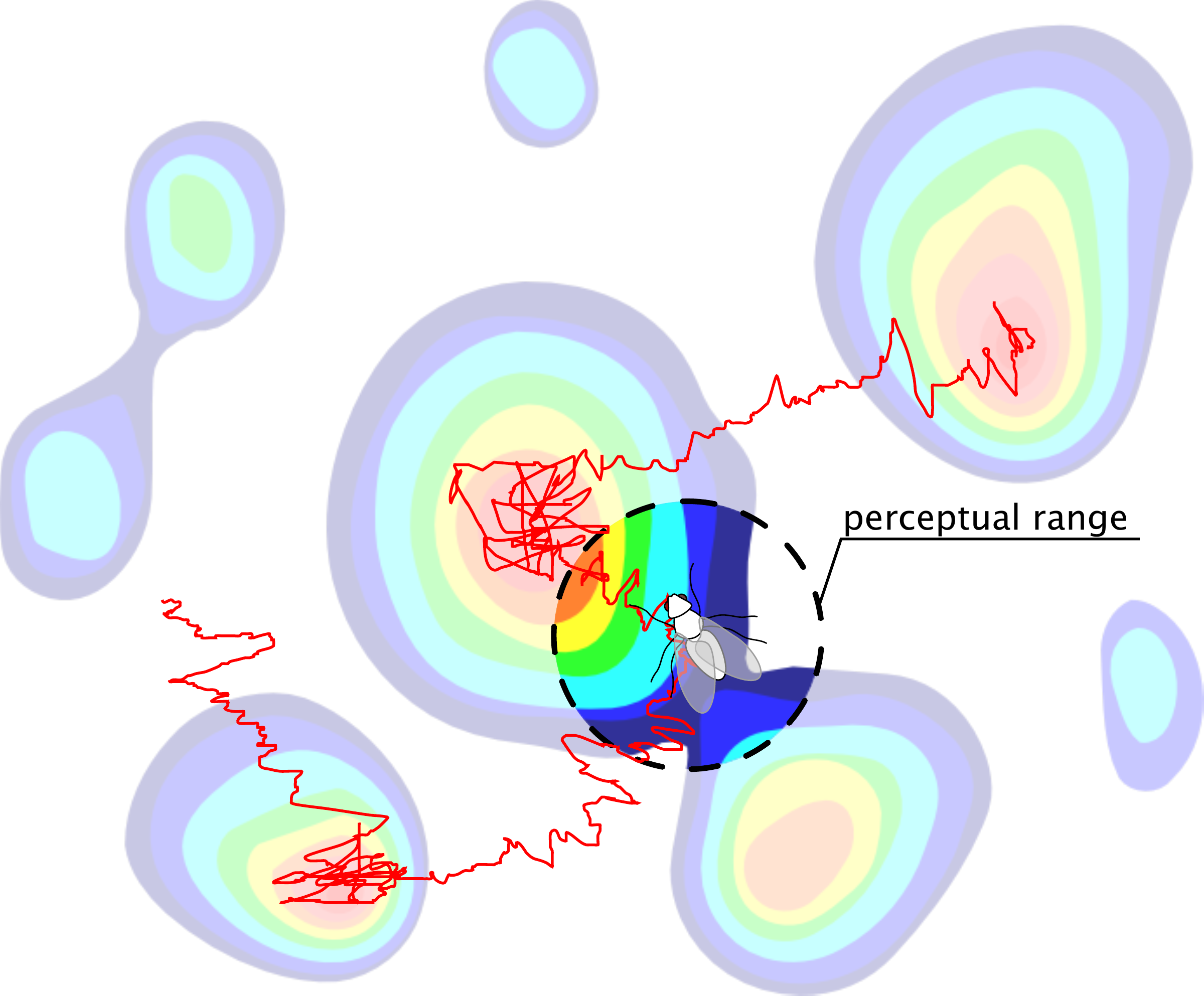}
	\caption{A searcher with a finite perceptual range navigating a heterogeneous landscape using a biased random walk search strategy. In contrast to standard local searchers,  which navigate based only on point-wise information, our searcher can use non-local information within its perceptual range to optimise its movement and exploration.}
	\label{Fig1}
\end{figure}

Exploration, movement, and search for resources are ubiquitous among organisms in nature~\cite{hein,bell,viswanathan}. Classical 
theories of search~\cite{stephens}, such as optimal foraging theory~\cite{macarthur,charnov}, have mostly focused on long time limits and typically assume that natural selection favours search strategies that maximise long-term encounters with nutrients. However, many phenomena in ecology~\cite{hastings} and other fields of biology operate in transient regimes~\cite{strelkowa,strelkowa2}, extending over time scales that never reach the asymptotic stationary state~\cite{schrodinger}. 
Another typical assumption is to consider random walks~\cite{berg:93,voituriez,benichou} or diffusion processes~\cite{okubo} to describe the movement of searchers navigating the landscape based on local information~\cite{vergassola,bartumeus,viswanathan3}. Yet, in many instances, searchers can obtain and store~\cite{moser} non-local information gathered through sensory cues~\cite{mistro,Gomez} or through anticipation of environmental changes~\cite{Mitchell,vandenbos,glnk} (Fig.~\ref{Fig1}). The question then arises as to how such finite perceptual range can influence both the dynamics of movement and the search efficiency over the finite time scales relevant in biology.

Here we study the role of finite time scales associated with ecological movement and search; specifically, the effect of limited spatial perception when the search time is itself finite. To formalise these aspects, we propose an \textit{optimal navigation} (ON) model, which allows us to extend the description of search as a biased random walk~\cite{codling,okubo,berg:93}, and reinterpret it in the framework of optimal control theory. 
The ON model includes a time horizon that quantifies the perceptual range of the searchers along their trajectory and fixes a non-local optimisation target for the agents. In the limit of vanishing time horizon (i.e., as the spatial perception shrinks and the information becomes local), the ON model recovers the classic Keller-Segel (KS) drift-diffusion model~\cite{KS} of \textit{local} search strategies (i.e, with instantaneous sensing and alignment to the point-wise gradient).

Using simulations and analytical results, we find that a population of non-local searchers moving towards a nutrient patch exhibits distinct transient behaviour, clustering faster at the hotspot than local searchers, thereby increasing their search efficiency. Our results show that the maximum efficiency gain occurs when the perceptual range of the searchers matches the environmental length scale over which the nutrient concentration changes significantly. As the search time becomes asymptotically large or small, the efficiency gain from the non-local strategy diminishes, and the searchers behave effectively as local responders. If the environmental length scale changes, we show that the efficiency gain can be maintained as long as searchers can adjust their sensitivity. This means that finite perception remains advantageous to searchers that can rescale their response dynamically, or to populations that contain a diversity of responses.
Finally, we consider the effect of interaction between searchers, and show that non-local information consistently reinforces the dominant strategy in the population and leads to improved search efficiency overall, even though multimodality (subpopulations) can appear during the transients. Our framework provides and optimisation perspective on a range of collective phenomena in population biology and, more generally, on biologically-inspired search and exploration algorithms, thus shedding light on the role of spatial perception on finite-time search.

\section{The classic Keller-Segel model: noisy search with local gradient alignment}   

A classic model for the dynamics of a population of searchers using local gradient alignment is given by the Keller-Segel equation, which we briefly recap here. 

Consider a population of searchers moving in a closed, bounded, $d$-dimensional domain $\Omega \subset\R^d$. The searchers move by responding to two concentration fields: to a primary stimulant $S_1(x)$ (e.g., nutrient) with sensitivity $\chi_1$, and to a secondary stimulant $S_2(x,t)$ (e.g., pheromone), released by the agents themselves, with sensitivity $\chi_2$.

On long space and time scales relative to the microscopic motion, one can describe the biased random walk $x(t) \in \Omega$ of searchers by a Langevin equation~\cite{codling}:
\begin{equation}
\frac{d x}{d t} = v(x, t) + \sqrt{2D} \,\xi( t)\,, \label{langevin2}
\end{equation}
where $ \xi(t) $ is a white noise process, $D$ is the coefficient of diffusion, and
\begin{equation}
	v(x, t) =  \nabla \left(\chi_1 \, S_1(x) + \chi_2 \, S_2(x,t) \right).   \label{v}
\end{equation}
is the \emph{velocity of the searcher}.
The parameters $ D,\,\chi_1$ and $\chi_2 $ are typically inferred experimentally from trajectories of the agents~\cite{oshanin,okubo}, and can sometimes be expressed in terms of microscopic parameters~\cite{othmer}. 
Note that Eqs.~\eqref{langevin2}--\eqref{v} describe a searcher that uses \emph{local information}, since it aligns its velocity instantaneously to the gradients of $S_1 $ and $S_2$. 

The secondary stimulant $ S_2 $ introduces interaction between the searchers. If $ S_2 $ is assumed to diffuse faster than the searchers, its evolution is given by
\begin{equation}
	S_2 (x,t) = \int_\Omega \Phi(x-x') \rho(x',t)\,dx' : = \Phi\ast \rho    \label{eq:S2_convol}
\end{equation}
where 
\begin{equation}\label{greens}
\Phi(x) = 
\begin{cases} 
-\dfrac{1}{2\pi} \log ||x||,  \quad & \text{for } d= 2 \\
\dfrac{\Gamma \left(\frac{d}{2}+1 \right)}{d(d-2)\pi^{d/2} } ||x||^{2-d} \quad & \text{for } d \neq 2, 
\end{cases}
\end{equation}
and $\Gamma $ is the Gamma function (see Appendix \ref{nondim}). 

Taking all together, the time evolution of the population density $\rho (x,t)$ of searchers obeying \eqref{langevin2}--
%\eqref{v} 
\eqref{eq:S2_convol} can be described with a Fokker-Planck equation~\cite{gardiner} known as the \textit{Keller-Segel} (KS) model. In dimensionless variables $ x \to x/L$ and $ t \to t/T_D$, where $T_D = L^2/D$ is the diffusion time of the searchers, the KS equation reads
\begin{equation}
	\partial_t\rho - \nabla^2\rho + \text{Pe}_1\nabla\cdot(\rho\nabla S_1) + \text{Pe}_2 \nabla \cdot \left( \rho\,\nabla (\Phi\ast\rho) \right) = 0,
	\label{eq:KS}
\end{equation}
where the parameters $ \text{Pe}_1 = \chi_1 S_\text{1,av}/D$ and $\text{Pe}_2 = \chi_2 S_\text{2,av}/D$ are 
P\'{e}clet numbers quantifying the ratio of diffusive to advective forces on the searchers, and $S_\text{1,av}$, $S_\text{2,av}$ are the average stimulant concentrations. 
Given an initial distribution $\rho_0(x) := \rho(x,0)$, a stimulant profile $S_1(x)$, and parameters $ \text{Pe}_1$ and  $\text{Pe}_2 $, Eq. \eqref{eq:KS} can be solved numerically using standard techniques.
% of finite volume schemes in the case of no interaction between searchers. 
Here, we use a first-order in time, second-order in space forward Euler scheme~\cite{CCH} with upwind discretisation and $ \Delta x = 0.01$ and $\Delta t = 10^{-6}$. 
We denote the solution of the KS model by $\rho_\text{KS}(x,t)$.
See Appendix~\ref{nondim} for a fuller derivation and details of the non-dimensionalisation. 

\subsection*{Variational rewriting of the KS model}

The KS model can be recast in a variational gradient formulation~\cite{carrillo3,jko}.
First, rearrange \eqref{eq:KS} as an advection equation: 
 \begin{align}
	\partial_t \rho &+ \nabla\cdot\left(\rho u \right)=0 \label{eq:advection}\\
     \text{with} \quad & u  = -\nabla \log\rho + v \,,\label{eq:velocity_advection}
\end{align}
where $ u $ is the \emph{velocity of the population}. 
Since the velocity of the searchers $v$ is a gradient~\eqref{v},  
$u$ is also a gradient:
\begin{equation}
\label{eq:u_gradient}
u  =  \nabla \left(-\log \rho + \text{Pe}_1 \, S_1 + \text{Pe}_2 \, \left( \Phi\ast\rho \right) \right).
\end{equation}
All terms in \eqref{eq:u_gradient} are either local or symmetric with respect to $ x $; hence $ u $ can be written~\cite{carrillo3} as a first variation 
\begin{align}
\label{eq:variation_functional}
u  = - \nabla \left(\frac{\delta\mathcal{F}}{\delta\rho}\right), 
\end{align}
where the \textit{free energy} functional
\begin{align}\label{fitfun}
	\mathcal{F}[\rho]=\int_\Omega \left( \rho \log \rho - \text{Pe}_1 \, \rho S_1 -  \frac{\text{Pe}_2}{2} \rho \, (\Phi \ast \rho ) \right) d x
\end{align}
includes (in order):  an entropic term from the stochastic component of the dynamics~\eqref{v}; the internal energy of the stimulant landscape $S_1$; and a term from the interaction between searchers via the secondary stimulant $ S_2 $. 

The KS model can then be rewritten in the equivalent gradient flow form
\begin{equation}\label{gradflow}
        \partial_t \rho  = \nabla\cdot\left(\rho \, \nabla \frac{\delta \mathcal{F}}{\delta \rho}\right),
\end{equation}
which has the important implication that the evolution of $ \rho(x,t) $ can be computed using the Jordan-Kinderlehrer-Otto (JKO) variational optimisation scheme~\cite{jko,bcc}. From an initial density $ \rho_0(x) $, the JKO scheme constructs 
a sequence of probability distributions $\{ \rho(x,k\Delta t)\}_{k \ge 0}$ 
\begin{equation}\label{JKO}
	\rho(x,(k+1)\Delta t)= \argmin_{\rho'}\left\{\frac{1}{2\Delta t} d_W (\rho(x,k\Delta t),\rho' ) + \mathcal{F}[\rho'] \right\},
\end{equation}
where $ \Delta t > 0 $ is the time step, and $ d_W(\cdot,\cdot) $ is the Wasserstein distance between two distributions. The solution~\eqref{JKO} has been proved to converge to the solution of Eq.~\eqref{eq:KS} in the limit $ \Delta t \to 0 $~\cite{bcc}. 

%%%%%%%%%%%%%%%%%%%%%%%%%%%%%%
%%%%%%%%%%%%%%%  Figure 2 
  \begin{figure*}[tbhp]
	\centering
	\includegraphics[width=\textwidth]{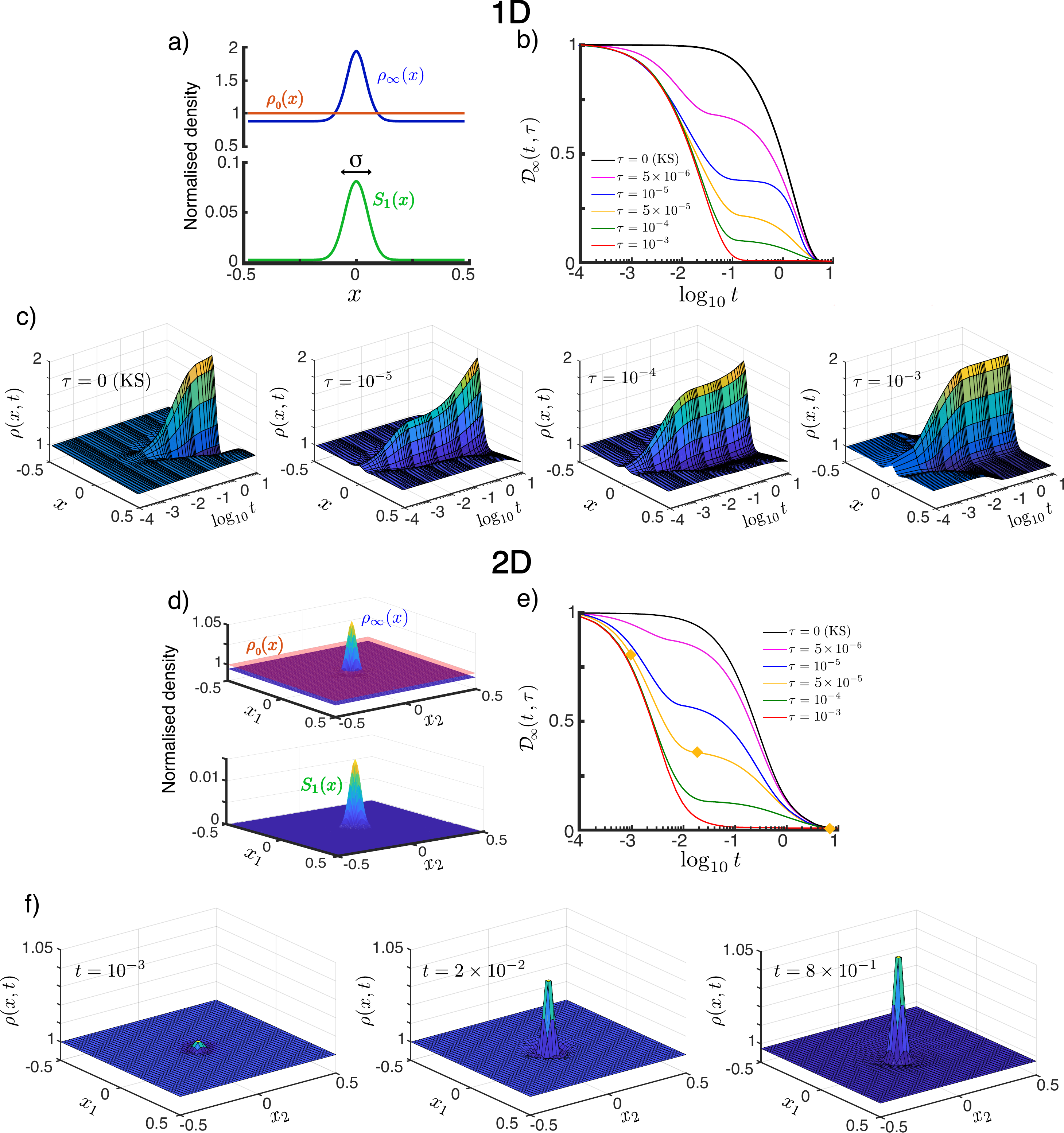}
	\caption{Transient population dynamics in nonlocal search with the ON model in the one-dimensional (a--c) and two-dimensional (d--f) cases.  	
	a) The searchers are initially uniformly distributed with $ \rho_0(x)=1 $. We simulate the time evolution of the population in a Gaussian stimulant profile $ S_1(x;\sigma) $~\eqref{gauss} until they reach the stationary state $ \rho_\infty(x)$~\eqref{eq:stationary}. 
	b) The approach towards stationarity measured by $\mathcal{D}_\infty(t; \tau)$~\eqref{D2}, the normalised $ L^2 $-distance of the solution to the stationary state. For small time horizons $ 0<\tau<\tau^*_\sigma $, there exists a long-lived intermediate transient state, whereas for $ \tau\ge\tau^*_\sigma $ the dynamics directly converges to the stationary state as the searchers quickly escape the diffusion-dominated part of the domain. See Fig.~\ref{Fig3} for the definition of $\tau^*_\sigma$.   
	c) Space-time plots illustrating the convergence towards the stationary state by the ON model in the one-dimensional case ($d=1$). The time evolution for increasing time horizon $ \tau = 0, 10^{-5}, 10^{-4}, 10^{-3} $ show qualitatively different transients. The KS model is equivalent to $\tau=0$. 
	d) - e) are equivalent to a) - b) but for the two-dimensional case. 
	f) Snapshots of the time evolution of the ON population with time horizon $\tau = 5 \times 10^{-5}$  in the two-dimensional case taken at increasing times ($t= 10^{-3}, 2 \times 10^{-2}, 8 \times 10^{-1}$, marked with yellow diamonds in e)). The convergence to stationarity displays the two-stage transient: the searchers near the centre aggregate during the initial fast transient, whereas the searchers far from the centre slowly diffuse towards the centre until the stationary state is reached. As the time horizon $\tau$ is increased 
	%, reduces (for $0<\tau<\tau^*_\sigma$) or eliminates (for $\tau>\tau^*_\sigma$)
	 this second slow transient dynamics is reduced. See animations of this solution in the supporting material. All simulations in this figure with $ \text{Pe}_1 = 2 $ and $ \sigma = 0.05 $.
	}
	\label{Fig2}
\end{figure*}
 %%%%%%%%%%%%%%%%%%%%%%%%%%%%%%

\section{The Optimal Navigation model: a population of searchers with non-local optimisation} 

The variational rewriting~\eqref{gradflow} and its approximation scheme~\eqref{JKO} leads us to formulate the \textit{optimal navigation} (ON) search model, as follows. Consider a population of searchers that move by performing the optimisation~\eqref{JKO} over a finite time horizon $ \tau \ge\Delta t> 0 $, which reflects the perceptual range of the agents. Then the time evolution of the population corresponds to a sequence of constrained optimisation problems~\cite{benamou}, i.e., a succession of JKO solutions, each over time $ \tau $. 

Starting from the initial density $ \rho_0(x) $, we construct the evolution of $ \rho(x,t) $, such that each iteration $ k \ge 0 $ finds 
$m(x,s):=\rho(x,k\tau + s)$ for $ s\in[0,\tau] $ by solving the minimisation problem:
\begin{equation}\label{ON}
    \begin{aligned}
    \underset{(m,u)}{\text{minimise}} \,\, & \mathcal{J}(m,u) := \int_0^{\tau}\int_\Omega m\frac{|u|^2}{2} dx ds + \mathcal{F}[m(x,\tau)]\\
    &\text{subject to} \,\; \partial_s m + \nabla\cdot(m u)=0.
    \end{aligned}
\end{equation}
Note that the constraint is the continuity equation ensuring the conservation of $ \rho $ as in~\eqref{eq:advection}, whereas the cost function $ \mathcal{J} $ contains a transportation cost, which constrains the average motion to geodesics between optimal states, and an end-point term involving the evaluation of the free energy at $\tau$~\eqref{fitfun}. Although we use it here for a particular form of the free energy functional, the formulation is generic: through suitable choice of $ \mathcal{F} $, the ON model~\eqref{ON} converges to a broad class of conservation laws as long as they can be recast as continuity equations and possess a variational structure~\cite{jko,carrillo3,bcc,CM}. 

We denote the solution of the ON model~\eqref{ON} by $\rho_\text{ON}(x,t;\tau)$, and compute it using Algorithm~\ref{alg:carola}, a gradient descent algorithm inspired by Ref.~\cite{carola} and presented in detail in Appendix~\ref{app:algo}.

Physically, the ON model~\eqref{ON} describes the motion of searchers that optimise their displacement over paths bounded by the time horizon $ \tau $. From the proof of the JKO scheme~\cite{bcc}, it follows directly that the ON model recovers the local KS model as $\tau \to 0$:
\begin{equation}
\lim_{\tau \to 0} \rho_\text{ON}(x,t;\tau) = \rho_\text{KS}(x,t)
\end{equation}
For finite horizon $ \tau > 0 $, the time evolution of the ON model departs from the KS solution due to the effect of non-local information on the movement of the searchers, as explored below.

\section{Non-local search: transients and enhanced search efficiency}   

We use the ON model~\eqref{ON} to study how the finite perception of the agents (encapsulated in the time horizon $\tau >0$) affects the search at the population level. We first consider non-interacting searchers insensitive to the secreted stimulant, i.e., $ \text{Pe}_2 = 0 $. The case of interacting searchers is presented in Section~\ref{sec:interaction}.

Our numerics start with a uniform initial condition $\rho_0(x)=1 $ and we compute $ \rho_\text{ON}(x,t;\tau)$, the time evolution~\eqref{ON} of the ON population of non-local searchers with time horizon $ \tau>0 $. We also compare it to the time evolution~\eqref{eq:KS} of a KS population of local searchers, or equivalently the ON model with $\tau = 0$.

Both the KS and ON models converge to the same stationary solution $\rho_\infty(x)$ asymptotically as $t \to \infty$:
\begin{align}
\label{eq:stationary}
	\rho_\infty(x) & : = \lim_{t\to\infty}\rho_\text{KS}(x,t) = \frac{1}{Z}\exp \left(\text{Pe}_1 \, S_1(x) \right) \\
	&=  \lim_{t\to\infty}\rho_\text{ON}(x,t; \tau),  \label{eq:stationary_ON}
\end{align}     
for all $ \tau $ and $S_1$, given by the Gibbs-Boltzmann distribution~\eqref{eq:stationary}, where $ Z $ is a normalisation constant. This result is well known for the KS equation~\cite{gardiner}. To check that $\rho_\infty(x)$ is also the stationary solution of the ON model, note that at stationarity $d_W(\rho_\infty,\rho')=0 $ in \eqref{JKO}, and the result~\eqref{eq:stationary_ON} follows from solving for the minimiser.

The approach to stationarity, on the other hand, reveals differences between the KS and ON models. To develop our intuition, let us first consider a linear attractant gradient $S_1(x) = \alpha x$. Such a landscape is fully characterised by the local gradient $\nabla S_1(x)$, and hence non-local searchers have no advantage since $\tau>0$ provides no further information than what is known by local searchers. In line with this expectation, when we solve the ON model \eqref{ON} using the method of characteristics (Section S2 in the SI), we find that the drift velocity predicted by the KS and ON models coincide for all $\tau$, i.e. $$u(x,t;\tau)\equiv u(x,t) \quad \text{when $S_1$ is linear}.$$  

This observation implies that non-local search is only advantageous in landscapes with non-zero curvature. To illustrate this point in more detail, let us consider a static Gaussian concentration of stimulant with characteristic length scale $ \sigma \ll 1$ over the domain $  \Omega = [-1/2,1/2]^d $ (Fig. \ref{Fig2}a):
\begin{equation}\label{gauss}
	S_1(x;\sigma) = \frac{1}{\sqrt{(2\pi)^d\text{det}(\Sigma)}} \exp\left(-\frac{1}{2}x^T\Sigma^{-1}x\right),
\end{equation}
where $\text{det}(\cdot)$ is the matrix determinant and $x^T$ is the transpose of the vector $x$. For simplicity of the computations that follow, we take $\Sigma = \sigma^2 I_d$, where $I_d$ is the $d$-dimensional identity matrix. Such a Gaussian landscape serves as a simple model of a stimulant patch emanating from a point source, and its characteristic length scale $\sigma$ indicates regions of steep attractant gradients near the source ($||x|| \ll \sigma$) and regions of shallow gradients in the tails of the distribution ($||x||\gg\sigma$).

In Fig.~\ref{Fig2} we show that, as the time horizon $\tau$ is increased, the population of ON searchers exhibits a faster approach to stationarity, as measured by the normalised $L^2$-distance between $\rho_\text{ON}(x,t;\tau)$ and $\rho_\infty (x)$ as a function of time for different values of $\tau$: 
\begin{align}
 \mathcal{D}_\infty(t; \tau)=\frac{|| \rho_\text{ON}(x,t;\tau)-\rho_\infty (x)||}{||\rho_0(x)-\rho_\infty (x)|| }.\label{D2}
\end{align} 
Fig.~\ref{Fig2}a-c presents the one-dimensional case ($d=1$). The effect of the horizon in accelerating the convergence to stationarity is shown in Fig.~\ref{Fig2}b. Note also that for small values of $\tau$, an intermediate, quasi-steady distribution develops during the transient (e.g, $\tau= 10^{-5}$ in Fig.~\ref{Fig2}b-c). This long-lived intermediate behaviour is the result of the population evolving on two timescales~\cite{okubo}: searchers near the maximum of $S_1(x)$ ($ ||x|| \ll  \sigma $) are driven by advection due to the steep gradient, whereas those far from the maximum ($ ||x|| \gg \sigma $) are driven by diffusion in shallow gradients, and hence move more slowly towards the maximum. Due to the slow diffusive searchers, the stationary state is only reached at a longer time scale $ t \sim 1$. As the horizon $\tau$ is increased, this dual behaviour (diffusion- or advection-dominated) is lost: the searchers escape quickly  the diffusion-dominated part of the domain and, as a result, the distribution approaches stationarity increasingly faster with no appreciable quasi-steady transient distribution (e.g, $\tau= 10^{-3}$ in Fig.~\ref{Fig2}b-c). 
The same behaviour is observed also in the two-dimensional case ($d=2$) in Fig.~\ref{Fig2}d-f.

%%%%%%%%%%%%%%%%%%%%%%%%%%%%%%
%%%%%%%%%%%%%%%  Figure 3
\begin{figure}
	\centering
	\includegraphics[width = \columnwidth]{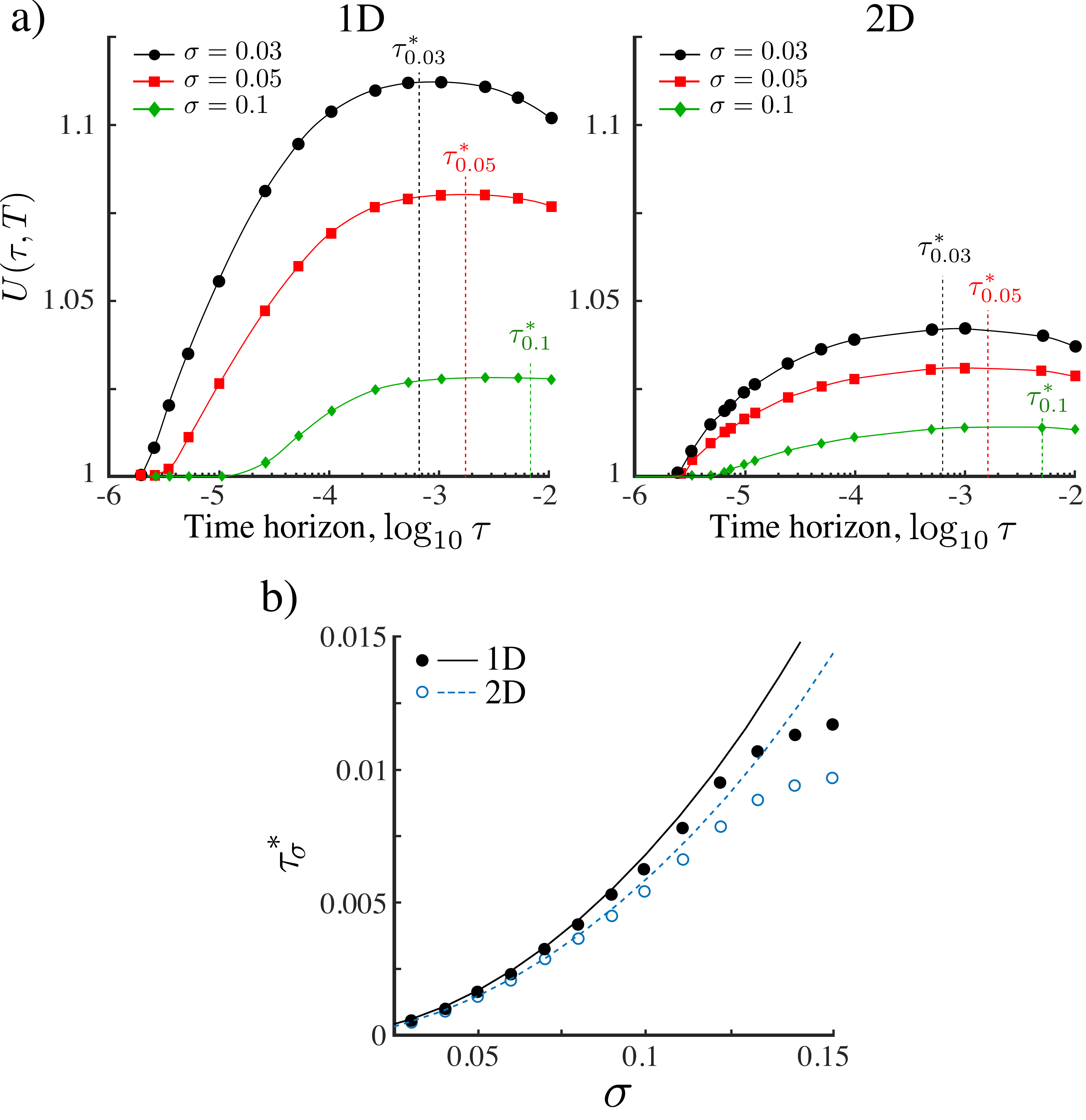}
	\caption{a) The relative search efficiency~\eqref{uptake} of the ON population in one and two dimensions as a function of the time horizon $\tau$ for three Gaussian stimulant profiles with length scales $ \sigma = 0.03,\,0.05,\,0.1$. All results are computed over a fixed search time $ T = 10^{-1} $. Values of  $U(\tau,T)>1$ indicate improved search efficiency of the ON population as compared to the KS population. The maximum efficiency $ U^*(\tau^*_\sigma,T) $ is achieved at a time horizon $ \tau^*_\sigma $. For a given $\text{Pe}_1$, the relative search efficiency decreases in higher dimensions due to the increasing dominance of diffusive motion relative to ballistic motion.
	b) Comparison of simulations (circles) with our estimate (line) of $ \tau^*_\sigma $ obtained from \eqref{msd}--\eqref{eq:S1} by matching the mean-squared distance travelled to the length scale of the landscape as given by Eq.~\eqref{eq:length_scale}. The estimate is accurate when~$ \sigma \ll 1 $.
	} 
	\label{Fig3}
\end{figure} 
%%%%%%%%%%%%%%%%%%%%%%%%%%%%%%

Such transient states can be important in biological systems, which typically operate on time scales far from the asymptotic long-time regime~\cite{hastings,strelkowa,strelkowa2}. In our setting, this situation arises when the \textit{search time} $T$ (which is analogous to the foraging effort in ecology) is smaller than the diffusion-dominated convergence time, i.e., when $T \ll 1$. In such a situation, non-local (ON) searchers have an advantage over local (KS) searchers since they converge faster to areas with high concentration of attractant.
To quantify this effect, we consider the amount of stimulant $S_1$ encountered over the search time~$T$
\begin{equation*}%\label{uptake_raw}
	\widehat{U}(\tau,T)=\int_0^T\int_\Omega S_1(x) \, \rho_\text{ON}(x,t;\tau)\, d x d t,
\end{equation*}
and define the relative \textit{search efficiency} as
\begin{equation}\label{uptake}
	U(\tau,T)=\frac{\widehat{U}(\tau,T)}{\widehat{U}(0,T)}.
\end{equation}
where $ \widehat{U}(0,T)$ is the uptake of the population of KS searchers.
Therefore, $ U(\tau,T) > 1 $ indicates a gain in search efficiency, that is, increased stimulant encountered by the population due to the perceptual horizon $ \tau>0 $.

Our numerics in Fig.~\ref{Fig3}a show that, given a finite search time $T$, the search efficiency~\eqref{uptake} reaches a maximum $U^*(\tau^*_\sigma,T)$ for searchers operating with an optimal horizon $\tau^*_\sigma$, which depends on the length scale $\sigma$ for a given dimension $d$ (Fig.~\ref{Fig3}a). 
The presence of a maximum follows from the asymptotic behaviour $U(\tau,T) \to 1$ for $\tau \to 0$ and $\tau \to \infty$. The latter limit  follows from the invariance of $\rho_\infty(x)$ under $\tau$, and the fact that the integral~\eqref{uptake} is asymptotically dominated by the steady state. The presence of a maximum is also observed in the two-dimensional case, but $U^*(\tau^*_\sigma,T)$ decreases in higher dimensions.

%%%%%%%%%%%%%%%%%%%%%%%%%%%%

\begin{figure*}[t!]
	\centering
    \includegraphics[width=.7\textwidth]{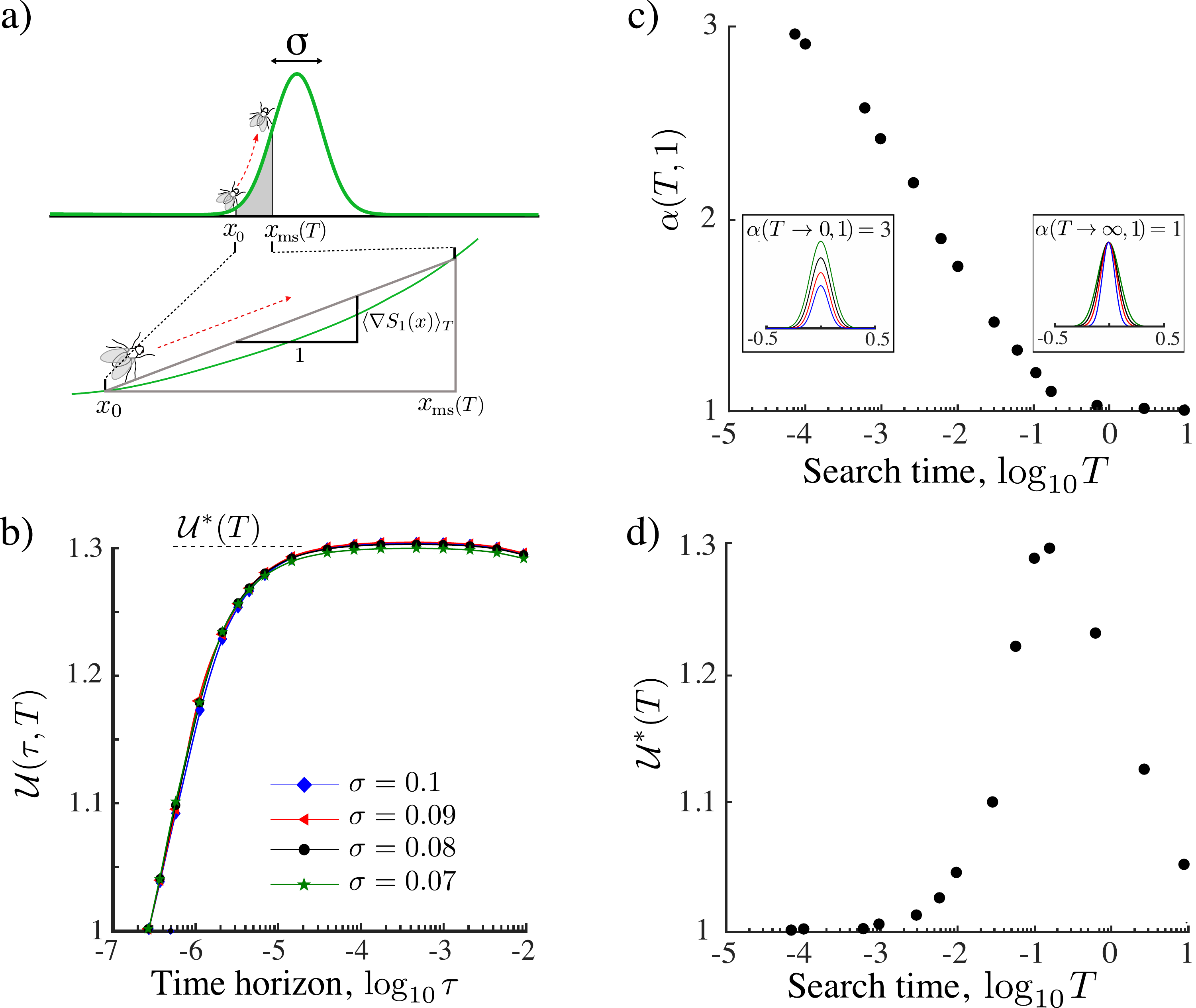}
    \caption{The search efficiency can be made invariant by scaling the sensitivity $\text{Pe}_1$. 
    a) The displacement of the searcher up to time $ T $ is approximated by the displacement in an effective gradient~\eqref{avegrad}. To account for the change in this effective gradient as $ T $ or $ \sigma $ varies, we renormalise the P\'{e}clet number as in~\eqref{scaledPe}. 
    b) The renormalisation yields invariance~\eqref{eq:invariance} of the search efficiency under $ \sigma $ ($ T=10^{-1} $ in this figure).
     c) The $ \alpha(T,d) $ for $d=1$ computed numerically matches the asymptotic results \eqref{alphaT}:  $ \alpha(T \to 0,1) =3 $ and $ \alpha(T \to \infty,1) =1 $. From the perspective of the searcher, the rescaling $\widetilde{\text{Pe}}_1 (\sigma, T)$ is equivalent to renormalising the landscape. Insets show examples of the renormalised Gaussian landscape with varying $ \sigma $ but the same effective gradient. 
     d) The optimal renormalised search efficiency $ \mathcal{U}^*(T) $ attainable by adapted agents achieves a maximum at search time $ T^* \sim 0.1 $. Hence non-local search is maximally advantageous for foraging times much smaller than the diffusion time $ T_D =1 $. 
      }
     \label{Fig4}
\end{figure*}

%%%%%%%%%%%%%%%%%%%%%%%%%%%%

The dependence of $\tau^*_\sigma$ with the length scale of the landscape $\sigma$ obtained numerically from our simulations is shown in Fig.~\ref{Fig3}b with solid (1D) and open (2D)  circles. To understand this dependence, consider a searcher at $x(t)$ obeying the Langevin equation~\eqref{langevin2} under the ON model. The reachable set until $ t+ \tau $ is within a ball of radius $x_\text{ms}(\tau)$ defined by
\begin{equation}\label{msd}
	x^2_\text{ms}(\tau) \simeq 2d\tau + \bar{v}^2(\tau) \tau^2,
\end{equation}
where the two terms represent the displacement due to diffusion and to an effective drift velocity $ \bar{v} $ over time $\tau$, respectively. To estimate $ \bar{v}(\tau) $, we assume that the search time $T$ is large enough such that individual searchers have explored the whole domain (e.g., for $ T=0.1 $ this is fulfilled when $ \tau>10^{-3} $ as seen in Fig. \ref{Fig2}b). 
The effective drift velocity can then be approximated by the velocity of an average searcher (over the domain) that maximises its gain up to time $ \tau $,
\begin{equation}	\label{eq:S1}
	\bar{v}(\tau)  \simeq  2 \text{Pe}_1  \int_{-1/2}^0\frac{S_1\left(x+x_\text{ms}(\tau)\right)-S_1(x)}{x_\text{ms}(\tau)} dx,
\end{equation}
where the integral is evaluated along a one-dimensional cross-section of the Gaussian landscape 
\begin{align*}
	\bar{v}(\tau)  &\simeq  \frac{ 2\text{Pe}_1 }{x_\text{ms}(\tau) } (2 \pi\sigma)^{-\frac{d}{2}} \int_{-1/2}^0 \left(
	e^{-\frac{(x+ x_\text{ms}(\tau))^2}{2 \sigma^2}} - e^{-\frac{x^2}{2 \sigma^2}}\right) dx\notag\\  
	& = \frac{\text{Pe}_1 }{x_\text{ms}(\tau)}(2 \pi\sigma)^{-\frac{d-1}{2}} \biggl[ \text{erf}\left( \frac{x_\text{ms}(\tau)}{\sqrt{2\sigma^2}}\right) - \text{erf}\left( \frac{1/2}{\sqrt{2\sigma^2} }\right) \notag\\
	&\quad \quad \quad - \text{erf}\left(\frac{x_\text{ms}(\tau) - 1/2}{\sqrt{2\sigma^2}}\right) \biggr]\,.
\end{align*}
From~\eqref{msd}~and~\eqref{eq:S1}, we obtain an estimate of the horizon $\tau$ necessary to search over a distance $x_\text{ms}$ with the ON model. 

The relevance of this estimate is shown in Fig.~\ref{Fig3}b, which shows that for small $\sigma$, the maximum search efficiency is attained when the mean-squared displacement of the searchers equals the length scale of the environment: 
\begin{align}
\label{eq:length_scale}
x_\text{ms} (\tau^*_\sigma) \simeq \sigma,
\end{align}
as obtained with our approximation. The ON model thus predicts that the most efficient searchers are those that tune their horizon such that they traverse the characteristic length scale of the environment within one optimisation step. Shorter or longer optimisations lead to a decreased search efficiency. 

The dependence of this behaviour on the dimension $d$ is also captured by \eqref{msd}, which tells us that the ballistic and diffusive terms balance when $\bar{v}(\tau) ^2\tau/(2d)$. Thus for a given $S_1$, $\tau$ and $\text{Pe}_1$ the motion of searchers becomes gradually diffusion dominated as the dimension $d$ increases. As a result, the relative search efficiency decreases in higher dimensions as shown in Fig.~\ref{Fig3}.

\section{Invariance of search efficiency through rescaling of response sensitivity}

The search efficiency of the ON model depends on the length scale $ \sigma $ of the Gaussian landscape: the ON gain diminishes as $\sigma$ increases (Fig.~\ref{Fig3}). However, as we now show, an ON searcher can retain the same search efficiency under a Gaussian landscape with varying length scale by adjusting $\text{Pe}_1$, its sensitivity to the stimulant.

To see this, consider an ON searcher starting at $ x_0 $ exposed to its nutrient micro-environment until time $T$. The effective gradient for this searcher depends on the starting position $ x_0 $ and is given by:
\begin{align}
\label{avegrad}
	\langle \nabla S_1 (x) \rangle_T  =   \frac{\int_{x_0}^{x_0+x_\text{ms}(T)} \left [ \frac{S_1\left(x+x_\text{ms}(\tau)\right)-S_1(x)}{x_\text{ms}(\tau)} \right ] dx}{x_\text{ms}(T)} .
\end{align}
For a fixed exploration time $T$, an increase in the stimulant length scale  $ \sigma $ leads to shallower effective gradients (Fig.~\ref{Fig4}a). Using asymptotic techniques, we show in Sect. S1 in the SI that the effective gradient~\eqref{avegrad} for symmetric Gaussian profiles has a well-defined behaviour in the two limiting regimes:
\begin{align}\label{alphaT}
\begin{cases}
\langle \nabla S_1 (x) \rangle_T  \propto  \sigma^{-2-d} &\text{ as } T\to 0, \\
\langle \nabla S_1 (x) \rangle_T  \propto  \sigma^{-d}  &\text{ as } T\to \infty.
\end{cases}
\end{align}
Together with the form of the dynamics~\eqref{langevin2}, this suggests the following scaling for the P\'eclet number:
\begin{equation}\label{scaledPe}
	\widetilde{\text{Pe}}_1 (\sigma, T) \propto \sigma^{\alpha(T,d)},
\end{equation}
with $d \leq \alpha(T,d) \leq 2+d $.

We have tested this scaling by obtaining the ON solution~\eqref{ON} over a given $T$ for Gaussian profiles with different $\sigma$ using the renormalised P\'eclet number~\eqref{scaledPe}. We then compute the relative search efficiency~\eqref{uptake} for this solution, $U(\tau, T;\, \widetilde{\text{Pe}}_1 (\sigma, T))$. Our numerics in Fig.~\ref{Fig4}b show that the search efficiency curves for the renormalised parameter~\eqref{scaledPe} for different $\sigma$ collapse on a single curve:
\begin{align}
\label{eq:invariance}
 U(\tau, T; \,\widetilde{\text{Pe}}_1 (\sigma, T)) \approx \mathcal{U}(\tau,T).
 \end{align}
The exponent $ \alpha(T,1) $, i.e. for $d=1$, is obtained numerically (Fig.~\ref{Fig4}c) is consistent with the expected asymptotic limits~\eqref{alphaT}. 
Note that, the effective gradient \eqref{avegrad} is a function of $x_\text{ms}(\tau)$, which changes with dimension. Therefore, although for any $d$ a scaling relationship exists, it will be different from the curve in Fig.~\ref{Fig4}c depending on $d$. 

Hence the search efficiency can be made invariant for different environmental length scales $ \sigma $ by rescaling the P\'eclet number~\eqref{scaledPe}. Alternatively, adjusting $ \alpha(T,d) $ can be viewed as responding to a 'renormalised landscape' $[\widetilde{\text{Pe}}_1 S_1 (x) ]$ in order to maintain the ON search efficiency. This is intuitive in limiting cases: when the search time is small ($ T \to 0 $), the efficiency remains unchanged on landscapes with similar local gradients near the centre ($ x_0 \ll 1 $); when the search time is large ($ T\to \infty $), the efficiency is invariant for landscapes with similar effective gradient over the whole domain (see inset of Fig. \ref{Fig4}c). 

This result suggests that searchers can optimise their search efficiency by adjusting their response sensitivity (as in the scaling~\eqref{scaledPe}) so as to balance the relative effect of the advection and diffusion velocities or, in other words, the relative importance of gradient optimisation versus noisy exploration. Since the diffusion coefficient $ D $ is typically independent of $S_1(x)$~\cite{othmer}, the adjustment of $ \text{Pe}_1 $ could be achieved by varying  the sensitivity as a function of the stimulant, i.e., $ \chi_1 (S_1(x)) $ (see Appendix~\ref{nondim}). In the Discussion, we explain possible biological mechanisms to achieve this effect.

Is non-local search advantageous over search times relevant for ecological systems?  The invariant search efficiency $\mathcal{U}(\tau,T)$ characterises the performance of a searcher that is tuned to the intrinsic length scale of the stimulant landscape during its search time $T$. In Fig. \ref{Fig4}d, we show the dependence of the maximum renormalised search efficiency~\eqref{eq:invariance} with the search time $T$. As expected, for short search time $ T \to 0$ and long search time $T \to \infty$, the efficiency $ \mathcal{U}^*(T) $ is equivalent to the local search strategy, i.e., $\mathcal{U}^*(T) \to 1$. 
However, between both extremes, searchers benefit from finite perception. Our numerics show that the optimal search time is $T^* = \argmax \,  \mathcal{U}^*(T, \tau^*)  \sim 0.1 < 1 = T_D$. 
Hence finite perception is maximally advantageous for search times smaller than the diffusion time, a fact that is typical in ecological systems (see Discussion).
  
%%%%%%%%%%%%%%%%%%%%%%%%%%%%%%%%%%%%%%%%

\section{Agent interaction in the Optimal Navigation model: increased efficiency and multimodality}
\label{sec:interaction}

%%%%%%%%%%%%%%%%%%%%%%%%%
\begin{figure*}[htp!]
\centering
\includegraphics[width=\textwidth]{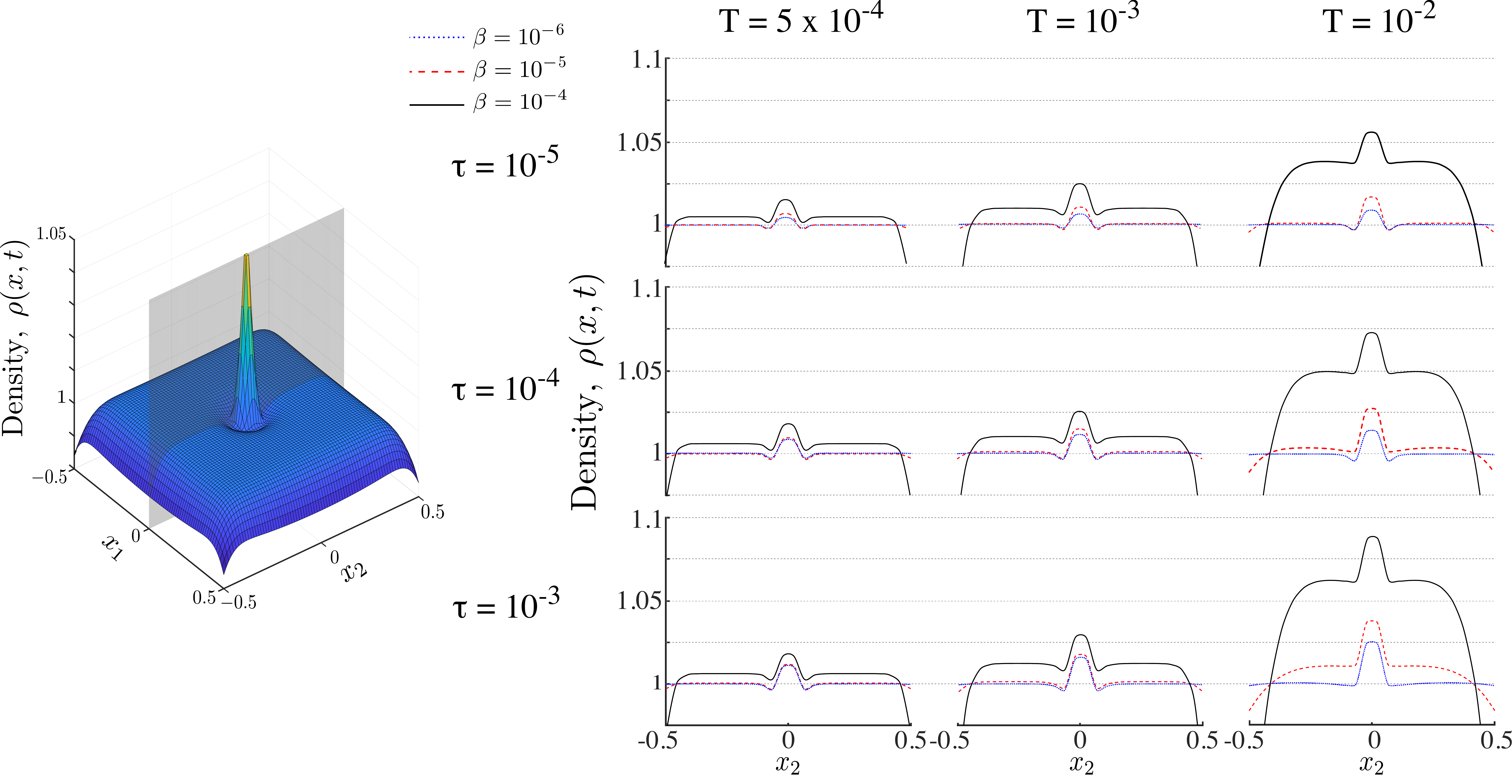}
\caption{Populations of interacting ON agents. Right panel shows snapshots of a population of ON searchers along the cutting plane shown in grey on the left hand panel. Snapshots are shown with time horizons $ \tau $ and interaction strength $ \beta $ for different search times $ T $. Interaction increases aggregation at the maximum of the nutrient profile (the centre of the domain) and reduces the long tails leading to the emergence of multimodality.
Simulations performed with $ \text{Pe}_1 = 2 $, $\sigma=0.05$.
}
\label{Fig5}
\end{figure*} 
%%%%%%%%%%%%%%%%%%%%%%%%%

\begin{figure*}[htp!]
\centering
\includegraphics[width=\textwidth]{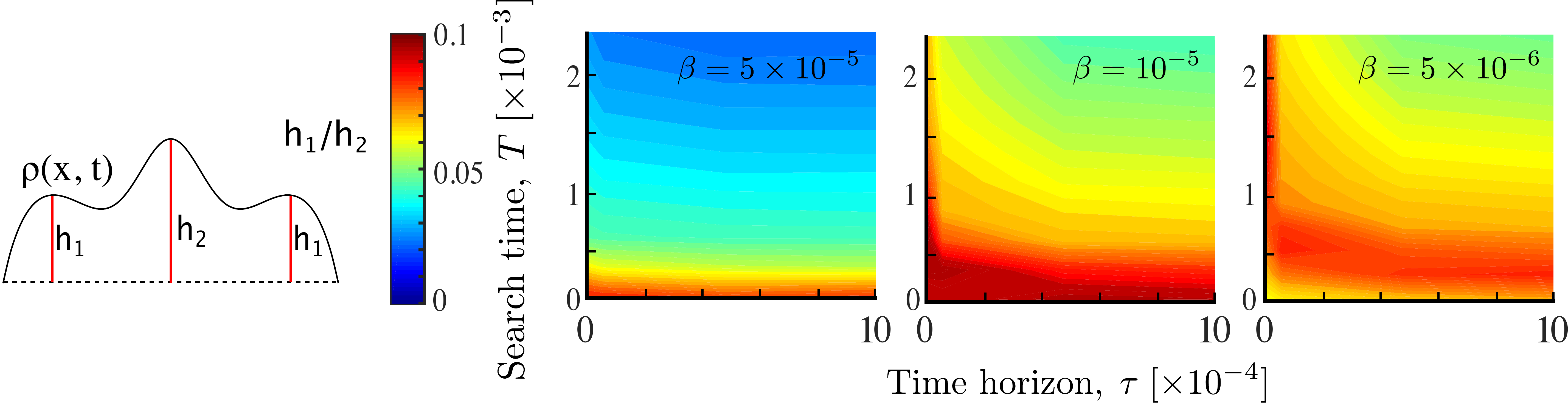}
\caption{Multimodality as a result of agent interaction.  The relative prominence of the smaller modes $h_1/h_2$ is computed for different parameters $ T $, $ \tau $ and $ \beta $. 
%When the interaction strength is higher ($\beta = 5\times 10^{-5}$) multimodality is present during the early transients. 
As the interaction $\beta$ decreases, multimodality gradually shifts to longer times and its magnitude also depends on the time horizon. This results from non-local searchers reaching more quickly a unimodal aggregate at the stimulant source. Simulations performed with $ \text{Pe}_1 = 2 $, $\sigma=0.05$. 
}
\label{Fig6}
\end{figure*} 
%%%%%%%%%%%%%%%%%%%%%%%%%

Until now, we have considered a population of non-interacting searchers reacting only to an external stimulant $S_1(x)$. Now we consider searchers which also interact with each other through an attractive secondary stimulant $ S_2 (x,t) $ (i.e., $ \text{Pe}_2>0 $). The effect of such interaction can be potentially contradictory: agent interaction may increase aggregation; however, aggregation might not increase search efficiency for the population if a large proportion of the agents clumps away from the stimulant source if the sensitivity to stimulant $ S_2 $ becomes larger. 

To explore these effects numerically, we use the ON model~\eqref{ON} with a weak interaction in the free energy \eqref{fitfun}
\begin{equation}
	\text{Pe}_2=\beta \, \text{Pe}_1,  \quad \beta \ll 1
\end{equation}
and compute the time evolution of the population for various $ \beta $ and $ \tau $. 
We restrict ourselves to the weak interaction case ($\beta \ll 1$) to facilitate our numerics. Specifically, to prevent infinite density concentration at finite time, a well-known artefact of the KS model, by applying a small regularising factor $\omega \rho^2$ with $0<\omega\ll 1$ to the free energy $\mathcal{F}[\rho]$~\eqref{fitfun}. This is a volume exclusion term that models the fact that agents occupy a finite volume in space~\cite{CC,carrillo2} (see \eqref{discrF} in Appendix~\ref{app:algo}). The weak interaction assumption ($\beta\ll 1$) is important for numerical performance %since increasing $\beta$ would result in higher velocities 
so as to avoid high velocities near the concentration point that would necessitate a finer time discretisation to maintain the CFL condition (see Appendix~\ref{app:algo}).

A summary of the numerics of the two-dimensional ON case with different horizons $\tau$ and levels of interaction $\beta$ over different search times $T$ is presented in Fig.~\ref{Fig5}. 
In all cases, the presence of interaction reduces the tails of the population density and increases aggregation near the maximum of $ S_1 $ at the centre of the domain. This becomes more noticeable as the search time grows and we approach stationarity (Fig. \ref{Fig5}, right column). During transients, however, agent interaction induces multimodality in the population (Fig. \ref{Fig5}, left and middle column). This implies that some searchers move away from the maximum of the stimulant $S_1(x)$ and aggregate into transient subpopulations. This behaviour arises due to the non-linear response of the searchers to the gradient of $S_1(x) $: for steep gradients of $S_1(x)$, agents are driven by attraction to $ S_1 $, whereas for shallow gradients of $S_1(x) $ the agents are driven by the interaction with the rest of the population through the secreted stimulant $ S_2 $. In contrast, for non-interacting agents ($\beta =0$) the distribution is always uni-modal (Fig. \ref{Fig5}). 

Fig. \ref{Fig6} shows the dependence of multimodality on $T$ and $\tau$ at three different interaction strengths $\beta = 5\times 10^{-5},\, 10^{-5},\,5 \times 10^{-6}$. We find that interaction introduces multimodality, as part of the population clumps away from the source of the attractant.  For higher interaction strength ($\beta = 5\times 10^{-5}$) multimodality is only present during early transients, and is independent of the time horizon $\tau$, since the strongly interacting population rapidly converges to the peak of the primary stimulant. As the interaction strength $\beta$ decreases, %($\beta =  10^{-5}$) 
multimodality lasts longer, but it decreases as the time horizon $\tau$ increases. This is in line with the expectation that increased spatial perception $ \tau $ or search time $T$ leads to overlapping information of the searchers about the environment, whereas, in contrast, when $\tau$ and/or $T$ are small, the searchers remain isolated within their local environment, thus leading to multimodality by local aggregation. 
%Finally, 
As the interaction decreases further 
%($\beta =  5\times 10^{-6}$) the magnitude of 
multimodality disappears.
% and, at the same time, the secondary modes appear at a slower timescale indicated by the shift in the maximum magnitude to higher search times $T$.

%Indeed, the optimal horizon in the present landscape is $\tau^*_{0.05} \sim 10^{-3}$ (Fig. \ref{Fig3}b) and correspondingly, the population remains unimodal at all times when $\tau\sim\tau^*_\sigma$ (Fig. \ref{Fig5}, bottom row). 

\begin{figure}
\centering
\includegraphics[width=.9\columnwidth]{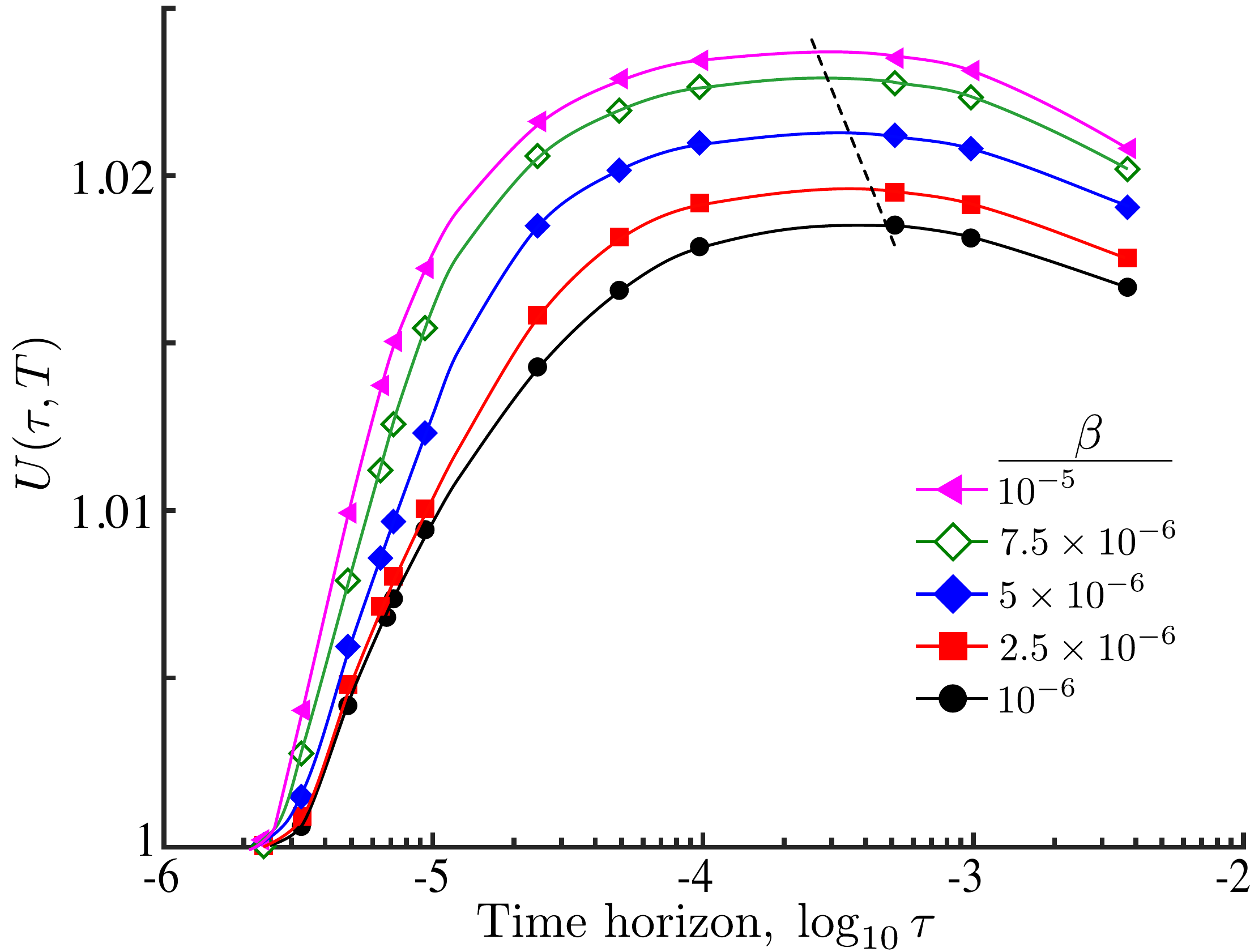}
\caption{Relative search efficiency~\eqref{uptake} of an ON population of interacting agents against time horizon with different interaction strengths $ \beta $. As $\beta$ grows, the search efficiency increases and the maximum is attained at smaller time horizons (dashed line). The improvement in the search efficiency with interaction follows from the increased accumulation of agents at the maximum of the nutrient and the reduced long tails far from the centre of the domain (Fig.~\ref{Fig5}). Simulation parameters: $ \text{Pe}_1=2,\,\sigma=0.05, \,  T = 5\times 10^{-2}$.  
}
\label{Fig7}
\end{figure}

Importantly, despite the presence of such mildly multimodal transients with clumped subpopulations, agent interaction $\beta$ leads to an overall improvement in the search efficiency $ U^*(\tau^*,T) $ at a shorter time horizon $\tau^*$ (Fig.~\ref{Fig7}). This behaviour follows from the increased concentration of searchers around the centre (on average), and suggest that when resources are sparsely distributed, agent interaction in conjunction with a finite perceptual range, could play a role in improving the collective sensing of the environment towards improved population efficiency.

\section{Discussion}

In this work, we studied how finite perception influences the dynamics and efficiency of collective random search in a population of agents. Using concepts from optimal control and random walks, we proposed a model that encapsulates the spatial information the searchers possess as a time horizon for an optimisation problem. Simulations of the dynamics of population search show that non-local information affects the movement strategy, as compared with the standard Keller-Segel model based on local optimisation. Although non-local search does not change the stationary state, it leads to qualitatively different transient responses of possible relevance in biological systems~\cite{hastings}. 
For example, marine bacteria have been observed to aggregate at food patches much faster ($ \sim10^1-10^2 $ s)~\cite{blackburn,seymour} than the timescale to reach the steady state distribution ($ T_D \sim 10^4-10^5 $ s, based on $ D \sim 10^2-10^3\;\mu\text{m}^2 $/s~\cite{kiorboe} and typical inter-patch distance $ L\sim 10^3\;\mu $m~\cite{azam}). Similarly, in several rodent species diffusion coefficients of $D\sim 200 $ m$^2/$day and home range of $L\sim 70 $ m have been reported~\cite{giuggioli}. Therefore, the time it takes to reach home by diffusion ($\sim 25$ days) is much longer than their typical response time ($ \sim 1 $ day).  

We find that non-local information can account for the increase in the search efficiency under transient search times: the maximum efficiency is reached when the mean-squared displacement of the searchers matches the environmental length scale of the stimulant. When the time horizon vanishes or when the search time is infinite, our model recovers the response of local searchers. This is in accordance with the fact that when long-range cues are unreliable, local response leads to highest efficiency~\cite{vergassola}. 

We also showed that the search efficiency can be made invariant to changes in environmental length scales by suitably scaling the response sensitivity. As a consequence, a searcher with a given perceptual range may always achieve its maximum efficiency by dynamically adjusting its sensitivity to the environmental stimuli. This can be achieved either by dynamically rescaling the responses via adaptation at the agent level~\cite{lazova}, or by the presence of a distribution of sensitivities among the agents at the population level. For example, it has been shown that phenotypic heterogeneity (or plasticity) across a population can be used to achieve maximum search efficiency in patchy environments~\cite{hollander}.

Finally, we considered the effect of interaction between searchers with finite perception and showed that interaction can lead to unimodal or multimodal population distributions on transient timescales. Multimodality appears even in the presence of unimodal stimulant landscapes due to a trade-off between following the environmental gradient or the rest of the population. In our numerics, interaction always improved the overall search efficiency of the population.

Recent theoretical and experimental studies \cite{mattar,pfeiffer} also suggest that rodents (and other higher animals) store spatial information of their environment, and in doing so commonly prefer longer (non-trivial) future paths as opposed to paths leading to immediate rewards. These results are in line with a predictive optimisation based on spatial knowledge, as in the ON model, where the searcher weighs up local cues with those at a distance to inform the planning of the future trajectory as opposed to an immediate (gradient) optimisation. 
Clearly, the perceptual horizon in the ON model is a very coarse abstraction of the spatial information the organism perceives (for example in its visual field) or is encoded in its neural or cellular memory. Going beyond our framework would require explicit assumptions about the sensory or cognitive mechanisms involved in search and navigation, an area of interesting future research which is out of the scope of this work.

Our work opens up several directions of research. Beyond our simple setup, it would be of interest to study search on temporally-fluctuating or patchy nutrient landscapes~\cite{redner} using non-local strategies. Random search theory based on local response predicts that when a searcher is positioned equally far from two nutrient patches it is equally likely to explore either patch. However, on transient timescales, nonlocal searchers are expected to explore the patch with denser resources with higher probability. This direction will be the object of future work.

\appendix
\section{The Keller-Segel model and its nondimensionalisation}\label{nondim}

On space and time scales much larger than the microscopic motion of the agents, we describe the biased random walk of the searchers with a Langevin equation \eqref{langevin2}-\eqref{v}, and the time evolution of the population density, $\rho (x,t)$, with the associated Fokker-Planck equation~\cite{gardiner}:
\begin{equation}
	\partial_{t}\rho -  D \, \nabla^2\rho + \nabla \cdot \left( \rho \,  \nabla \left(\chi_1 S_1 + \chi_2 S_2 \right) \right) = 0.
	\label{eq:KS_dim}
\end{equation}
If the agents release the stimulant $S_2$ at a constant rate $\Gamma $ (in units of mol/(agent t)) and $S_2$ diffuses with diffusion coefficient $D_2$, we obtain the forced diffusion equation:
\begin{equation}
	\partial_t S_2 - D_2 \nabla^2 S_2 = \Gamma\rho. \label{S2dim}
\end{equation}

To reduce the number of parameters, we non-dimensionalise the equations. We use the length scale $L$ and time scale $ T_D = L^2/D $ to define%  \MB{Any issues with 1D here?  $L$ ?  Does it have to be a volume?}
\begin{gather*}
 \hat{x} := x/L \quad \hat{t} := t/T_D.
\end{gather*}
Let $\hat{S}_2 := S_2/S_\text{2,av}$ and $\hat{\rho} := \rho/\rho_\text{av}$ with $\rho_\text{av} = \frac{1}{L^d} \int_\Omega \rho(x) dx$ and $S_\text{2,av} = \Gamma L^2 \rho_\text{av}/D_2$. Then Eq.~\eqref{S2dim} can be rewritten in non-dimensional form:
\begin{equation}
	\frac{D}{D_2}\, \partial_{\hat{t}} \hat{S}_2 - \nabla_{\hat{x}}^2 \hat{S}_2=  \hat{\rho}. 	
	\label{S2nondim}
\end{equation}
If $ S_2 $ diffuses faster than the searchers ($   D/D_2  \ll 1 $), the time-dependence in \eqref{S2nondim} may be neglected to first order and the resulting Poisson equation has the quasi-steady solution:
\begin{equation}\label{S2}
	\hat{S}_2 (\hat{x}) = \int_\Omega \Phi(\hat{x}-x') \hat{\rho}(x',t)\,dx' : = \Phi\ast \hat{\rho},
\end{equation}
where $ \Phi(x) $, given by \eqref{greens}, is the Green's function solution to the Poisson equation in $d$ dimensions.

Defining $\hat{S}_1 := S_1/S_\text{1,av}$ with $S_\text{1,av} =  \frac{1}{L^d} \int_\Omega S_1(x) dx$ we rewrite Eq.~\eqref{eq:KS_dim} in non-dimensional form:
\begin{equation}
	\partial_{\hat{t}} \hat{\rho} - \nabla_{\hat{x}}^2\hat{\rho} +  \nabla_{\hat{x}}\cdot \left( \hat{\rho} \, \nabla_{\hat{x}} \left( \text{Pe}_1 \hat{S}_1 + \text{Pe}_2 \hat{S}_2 \right)  \right) = 0,
	\label{eq:KS_nondim}
\end{equation}
where 
$\text{Pe}_1 =  \chi_1 S_\text{1,av} /D$ and $ \text{Pe}_2 =  \chi_2 S_\text{2,av}/D $ are P\'{e}clet numbers.
To simplify the notation, we drop the hats in all variables in the main body of the paper and always consider non-dimensionalised equations. From~\eqref{eq:KS_nondim}~and~\eqref{S2}, we get the non-dimensional \textit{Keller-Segel} model~\eqref{eq:KS}.

\section{Algorithm and numerical scheme for the solution of the Optimal Navigation model}
\label{app:algo}
%%%%%%%%%%%

%%%%%%%%%%%%
\begin{algorithm}
\caption{Gradient descent algorithm for \eqref{ON} based on Ref.~\cite{carola}. Simulation parameters used: $ \Delta x = 0.01 $, $ \Delta t = 10^{-6} $, $ \kappa = 1$, $ \delta = 10^{-6} $.}
\begin{algorithmic}
\\\\ \hrule \\
\STATE{\textbf{Input } $ \rho(x,k\tau) $, $ \kappa $, $ \delta $}\\
\STATE{\textbf{Output } $ m(x,s) := \rho(x,k\tau + s),\; s\in[0,\tau] $,}\\

\STATE{Set $ m^0_0 \gets \rho(x,k\tau) $}
\WHILE{$ ||u^{p+1}-u^p||_{L^2}>\delta $}\\
	\STATE{\textit{Compute the FP equation \eqref{fp} forward in time}}
    \FOR{$j=0:1:N$}
        \STATE{Compute $ m^{p+1}_{j+1} = (I - B(u_j^p))m^p_j $}
        \STATE{Set $ u^{p+1} \gets m^{p+1}_N$ and $ \phi^{p+1}_N \gets -\partial \mathcal{J}_{\Delta t\Delta x}( u^{p+1}) $}
    \ENDFOR \\
    \STATE{\textit{Compute the HJ equation \eqref{hjb} backward in time}}
    \FOR{$j=N+1:-1:1$}
        \STATE{Compute $ \phi^{p+1}_{j} = (I-B^T(u^p_j))\phi^p_{j+1} - \frac{\Delta t}{2}r^p_j $}
    \ENDFOR \\
    \STATE{Check $ [\, \mathcal{J}_{\Delta t, \Delta x}(m^p,u^p)-\mathcal{J}_{\Delta t, \Delta x}(m^{p+1},u^p) \,]< 0 $}
    \STATE{Update $ (1+\tau)u^{p+1} = \nabla \phi^{p+1} + \kappa u^p$}
    \STATE{Set $ p \gets p+1 $}
\ENDWHILE

\STATE{\textbf{Return } $ m $\\\\}
\hrule
\end{algorithmic}
\label{alg:carola}
\end{algorithm}
%%%%%%%%%%%%

For notational simplicity we present the numerical solution of the ON model in the one-dimensional case. The extension to two dimensions is straightforward. The ON model \eqref{ON} is numerically solved as an implicit forward Euler scheme, where we obtain during each step the evolution $ m(x,s)=\rho(x,k\tau + s) $ for the time window $ s\in [0,\tau] $ using a recently developed gradient descent algorithm~\cite{carola}. We used this algorithm since it preserves a discrete analogue of the variational structure of the ON model and has the required property that the free energy \eqref{fitfun} decays along the time evolution of $ m(x,s) $.

%The free energy \eqref{fitfun} decays in time along the time evolution of both the KS and the ON models. Preserving the variational structure at the numerical approximation is desirable to maintain a discrete analog of the decrease of the free energy \eqref{fitfun}. The numerical scheme used \cite{CCH} has this property due to the finite volume approach via upwinding and centered finite differences discretisation of the velocity field.

We first derive the optimality conditions of \eqref{ON}. The corresponding Lagrangian is~\cite{benamou}:
\begin{equation}
	\begin{aligned}
	&\mathcal{L}^\tau(m,u,\phi):=\mathcal{J}(u,m) \\ 
	&+ \int_0^\tau\left[\mathcal{F}[m(x,\tau)]- \int_\Omega\left( \partial_s\phi +  u\cdot \nabla\phi\right)m \,dx\right]ds \\
	&+ \int_\Omega \left[\phi(x,\tau)m(x,\tau)-\phi(x,0) \right ] m(x,0)dx\,,
	\end{aligned}
\end{equation}
allowing us to write \eqref{ON} as a saddle node problem:
\begin{equation}\label{saddle2}
	\inf_{(m,u)}\sup_\phi \mathcal{L}^\tau(m,u,\phi),
\end{equation}
where the infimum is taken over all the possible time evolutions of the population, i.e. pairs $ (m,u) $  satisfying $ \partial_s m + \nabla\cdot(m u) = 0 $. Solving for the first-order optimality conditions of \eqref{saddle2},
\begin{equation*}
	\partial_m \mathcal{L}^\tau = 0,\quad \partial_u \mathcal{L}^\tau = 0,\quad \partial_\phi \mathcal{L}^\tau = 0,
\end{equation*}
yields a variational mean-field game system~\cite{ll1,hcm} consisting of the Fokker-Planck equation~\eqref{fp} solved forward in time coupled to a Hamilton-Jacobi (HJ) equation~\eqref{hjb} solved backward in time:
\begin{subequations}\label{varMFG}
    \begin{align}
      &  \partial_s m +\nabla\cdot{(m u)}=0,\,\, m(\cdot,0)=\rho(\cdot,k\tau)\label{fp}\\
       & \partial_s\phi + u\cdot\nabla\phi+\frac{|u|^2}{2}=0,\,\, \phi(\cdot,\tau)=-\frac{\delta\mathcal{F}[m]}{\delta m}\biggr|_{s=\tau},\label{hjb}
    \end{align}
\end{subequations}
with $ u =-\nabla \phi $. The terminal condition in \eqref{hjb} can be explicitly solved as
\begin{align*}
	\frac{\delta\mathcal{F}[m]}{\delta m} &= \log m(x,t) - \text{Pe}_1 S(x) + \text{Pe}_2 \log|x|\ast m\,.
\end{align*}

Consider a uniform discretisation of $ \Omega = [-1/2,1/2]\times [0,\tau]$ as a grid $ (i,j) $, $ i=0,\dots, M+1 $, $ j=0,\dots,N+1 $ with steps $ \Delta x = 1/(M+1) $ in space and $ \Delta t = \tau/(N+1) $ in time for positive integers $ M,N $, where the first and last points are boundary nodes. Let $ m_{i,j} := m(i\Delta x,j\Delta t) $ on the grid points and $ u_{i+1/2,j} := u( (i+1/2)\Delta x, j\Delta t) $ on the staggered grid points. Finally, we impose Neumann conditions at the boundaries, $u_{1/2} = u_{M+3/2} = 0$, and denote the array of point excluding the two boundary nodes by $ u_j = (u_{i+1/2,j})_{i=1,\dots,M} $ and $ m_j = (m_{i,j})_{i=1,\dots,M} $.

We discretise the cost function $ \mathcal{J}[m,u] $ as
{\small
\begin{align}
	& \mathcal{J}_{\Delta t, \Delta x}(m,u):= \Delta x \Delta t \sum_{i=1}^{M} \sum_{j=0}^{N}\left( \frac{1}{2}r_{i,j}m_{i,j} + f_i(m_N)m_{i,N} \right), 
\end{align}}
where $r_{i,j} = ((u_{i-1/2,j})^2 + (u_{i+1/2,j})^2)/2 $,
{\small
\begin{equation} \label{discrF}
\begin{aligned}
f_i(m_j) &= \log(m_i)-\text{Pe}_1 S  
+ \text{Pe}_2 \sum_{l=1}^M\log(|(i - l)\Delta x|+\epsilon)m_{l,j} \\
&+ \omega m_{i,j}^2.
\end{aligned}
\end{equation}
}
Here $ \epsilon =10^{-5} $ and $\omega= 10^{-6}$ are small regularising factors to bound the logarithmic term and to guarantee the existence of finite solutions at all times~\cite{CC,carrillo2}.

We use a first order upwind scheme for the advection in \eqref{fp}:
{\small
\begin{align*}
	m_{i,j+1} &= m_{i,j} + \frac{\Delta t}{\Delta x}\left(m_{i+1/2,j}u_{i+1/2,j} - m_{i-1/2,j}u_{i-1/2,j} \right), 
\end{align*}
}	
\begin{align*}
& \text{where} \quad 	m_{i+1/2,j} = 
	\begin{cases}
		m_{i+1,j}, &\mbox{if } u_{i+1/2,j}<0 \\
		m_{i,j}, &\mbox{if } u_{i+1/2,j}\ge 0.
	\end{cases}  \\
        u_{1/2,j} &=u_{M+1/2,j}=0,  j = 0,\dots,N  \quad \text{(Neumann BC's)}.
\end{align*}
In compact form this may be written as 
\begin{equation} \label{discfp}
	m_{j+1} = (I - B(u_j))m_j,
\end{equation}
where $ I \in \mathbb{R}^{M\times M} $ is the identity matrix and $ B(u) \in \mathbb{R}^{M\times M} $ is the matrix associated with the upwind scheme. 

The corresponding scheme for \eqref{hjb} is given by 
\begin{align}\label{dischjb}
	\phi_{j} = (I - B^T(u_j))\phi_{j+1} + \frac{\Delta t}{2}r_j.
\end{align}

Algorithm~\eqref{alg:carola} may then be used with a given descent step $ \kappa $ to compute $ m(x,s) $ to the required tolerance level $ \delta $. Note that $ \kappa $, $ \Delta t $ and $ \Delta x $ must be chosen so as to fulfil the CFL condition~\cite{leveque}: $ u_{i+1/2,j}\Delta t /\Delta x \le 1 $ for all staggered grid points $ (i+1/2,j) $.  Then the cost function $ \mathcal{J}_{\Delta t, \Delta x} $ decreases at every iteration $ p $.

\section*{Acknowledgements}
AG acknowledges funding through a PhD studentship under the BBSRC DTP at Imperial College (BB/M011178/1). JAC acknowledges funding from the EPSRC EP/P031587/1. MB acknowledges funding from the EPSRC (EP/N014529/1 and EP/I032223/1).

\bibliography{refs}
\newpage
\onecolumngrid
\unappendix
\setcounter{figure}{0}
\setcounter{section}{0}
\setcounter{equation}{0}
\setcounter{table}{0}

\renewcommand\thesection{S\arabic{section}}
\renewcommand\thefigure{S\arabic{figure}}
\renewcommand\theequation{S\arabic{equation}}
\renewcommand\thetable{S\arabic{table}}

\begin{center}
{\LARGE Supplementary information: \\ Collective search with finite perception: transient dynamics and search efficiency}
\end{center}

\author{Adam Gosztolai}
\affiliation{Department of Mathematics, Imperial College London, London, United Kingdom}
\author{Jose A. Carrillo}
\affiliation{Department of Mathematics, Imperial College London, London, United Kingdom}
\author{Mauricio Barahona}
\affiliation{Department of Mathematics, Imperial College London, London, United Kingdom}

\date{\today}

\maketitle

\renewcommand\thesection{S\arabic{section}}
\renewcommand\thefigure{S\arabic{figure}}
\renewcommand\theequation{S\arabic{equation}}
\renewcommand\thetable{S\arabic{table}}

\section{Computation of the effective gradient under limited spatial exploration}

Here, we compute the effective gradient of a searcher, while relaxing the assumption that the searchers have explored the whole domain. In this case, the effective gradient of a searcher starting at $ x_0 $, will be given by the average gradient in the explored domain up to the search time $ T $. In one dimension, this is
\begin{align}
	\langle\nabla S (x)\rangle_T &=  \frac{1}{x_\text{ms}(T)}\int_{x_0}^{x_0+x_\text{ms}(T)}\frac{S_1\left(x+x_\text{ms}(\tau)\right)-S_1(x)}{x_\text{ms}(\tau)} dx\notag\\
	&= \frac{ A}{x_\text{ms}(T)x_\text{ms}(\tau)} \int_{x_0}^{x_0+x_\text{ms}(T)} \exp\left(-\frac{x^2}{2\sigma^2}\right)\left[\exp\left(-\frac{2xx_\text{ms}(\tau) + x_\text{ms}(\tau)^2}{2\sigma^2}\right) -1 \right]dx,\label{int1}
	%&= \frac{ A}{x_\text{ms}(T)x_\text{ms}(\tau)} \int_{-x_\text{ms}(T)}^{0} \exp\left(-\frac{(x+x_0+x_\text{ms}(T))^2}{2\sigma^2}\right)\left[\exp\left(-\frac{2(x+x_0+x_\text{ms}(T))x_\text{ms}(\tau) }{2\sigma^2}\right) -1 \right]dx.
	%&\approx \frac{ A}{x_\text{ms}(T)x_\text{ms}(\tau)} \int_{x_0}^{x_0+x_\text{ms}(T)} \exp\left(-\frac{x^2}{2\sigma^2}\right)\left[\exp\left(-\frac{xx_\text{ms}(\tau) }{\sigma^2}\right) -1 \right]dx 
\end{align}
where $A = 1/\sqrt{2\pi \sigma^2}$ is a normalisation constant.

Then, we use the assumption $ x_\text{ms}(\tau)\ll x_\text{ms}(T) $, and evaluate \eqref{int1} asymptotically to first order in two extreme cases: when $ x_\text{ms}(T)\ll \sigma $ and $ \sigma\ll x_\text{ms}(T) $. First, for limited search times we have that $ x_\text{ms}(\tau)\ll x_\text{ms}(T)\ll \sigma \ll 1 $. Therefore, we may write 
\begin{align}
	\langle\nabla S (x)\rangle_T &= \frac{ A}{x_\text{ms}(T)} \int_{x_0}^{x_0+x_\text{ms}(T)} \exp\left(-\frac{x^2}{2\sigma^2}\right) \left[ -\frac{x }{\sigma^2} +\mathcal{O}\left(\frac{x_\text{ms}(\tau) }{\sigma^2}\right) \right]dx\notag\\
	&= \frac{ A}{x_\text{ms}(T)} \int_{x_0}^{x_0+x_\text{ms}(T)} \exp\left(-\frac{x^2}{2\sigma^2}\right) \left[ -\frac{x }{\sigma^2} \right]dx +\mathcal{O}\left(\frac{x_\text{ms}(\tau) }{\sigma^2}\right)\notag\\
	&= \frac{ A}{x_\text{ms}(T)} \left[ \exp\left(-\frac{(x_0+x_\text{ms}(T))^2}{2\sigma^2}\right) -  \exp\left(-\frac{x_0^2}{2\sigma^2}\right)\right]+\mathcal{O}\left(\frac{x_\text{ms}(\tau) }{\sigma^2}\right).\label{int2}
\end{align}
Therefore, when $ x_0\sim 1 $, we obtain that $ \langle\nabla S (x)\rangle_T =  \mathcal{O}\left(x_\text{ms}(\tau) / \sigma^2 \right) $, meaning that for cells far from the origin the contribution to motion by gradient-driven advection can be neglected on short time scales. Meanwhile, close to the origin, $ x_0\approx 0 $, we may expand the argument in the exponential in \eqref{int2}, namely
\begin{align}
	\langle\nabla S (x)\rangle_T & = \frac{ A}{x_\text{ms}(T)} \left[ -\frac{(x_0+x_\text{ms}(T))^2}{2\sigma^2} +\frac{x_0^2}{2\sigma^2} \right]+\mathcal{O}\left(\frac{x_\text{ms}(\tau) }{\sigma^2} \right)\notag\\
	& = \frac{Ax_\text{ms}(T)}{\sigma^2} + \mathcal{O}\left(\frac{x_\text{ms}(\tau) }{\sigma^2} \right) + \mathcal{O}\left( \frac{x_0}{\sigma^2} \right)\notag\\
	& \propto  \frac{1}{\sigma^3} + \mathcal{O}\left(\frac{x_\text{ms}(\tau) }{\sigma^2} \right) + \mathcal{O}\left( \frac{x_0}{\sigma^2} \right). \label{int3}
\end{align} 

Second, for large search times, the search behaviour is independent of the starting point thus we may set $ x_0=0 $ in \eqref{int1}. If in addition, we have that $ x_\text{ms}(\tau) \ll \sigma \ll x_\text{ms}(T) \sim  1 $, then the first order asymptotic solution to \eqref{int1} reads:
\begin{align}
	\langle\nabla S (x)\rangle_T & = \frac{ A}{x_\text{ms}(T)}  \int_{0}^{x_\text{ms}(T)} \exp\left(-\frac{x^2}{2\sigma^2}\right)\left[-\frac{x }{\sigma^2}  \right] dx +\mathcal{O}\left(\frac{x_\text{ms}(\tau) }{\sigma^2}\right) \notag\\
	& =  -\frac{ A}{x_\text{ms}(T)} \left[ \exp\left(-\frac{x_\text{ms}^2(T)}{2\sigma^2}\right) -  1\right]+\mathcal{O}\left(\frac{x_\text{ms}(\tau) }{\sigma^2}\right)\notag\\
	&\propto \frac{1}{\sigma}+\mathcal{O}\left(\frac{x_\text{ms}(\tau) }{\sigma^2} \right) + \mathcal{O}\left(\frac{\sigma}{x_\text{ms}(T)} \right).
\end{align}
If on the other hand $ \sigma \ll x_\text{ms}(\tau)\ll x_\text{ms}(T) \sim  1 $, then we get that
\begin{align}
	\langle\nabla S (x)\rangle_T &=  2 A \biggr[\left(1-e^{-x_\text{ms}(T)/(2\sigma^2)}\right) - \frac{x_\text{ms}(\tau) }{4\sigma}e^{-x_\text{ms}(T)/(2\sigma^2)} \biggr] + \mathcal{O}(x_\text{ms}^2(\tau))\notag\\
	&\propto \frac{1}{\sigma} + \mathcal{O}\left(\frac{x_\text{ms}^2(\tau)}{\sigma^2}\right) + \mathcal{O}\left(\frac{\sigma}{x_\text{ms}(T)} \right).
\end{align}
Thus, to first order, when $ x_\text{ms}(T)\ll \sigma $ we have that $ \langle\nabla S (x)\rangle_T\propto \sigma^{-3} $, while for $ \sigma\ll x_\text{ms}(T) $ we have that $ \langle\nabla S (x)\rangle_T\propto \sigma^{-1} $.

%Finally, we consider the case, when the search time scale is commensurate with the time scale of environmental variation, i.e. $ x_\text{ms}(\tau)\ll x_\text{ms}(T) \sim \text{ord}( \sigma) $. In this regime, we may split the range of integration at 
%
%\begin{align}
%	\nabla_T S (x;\sigma) &= \frac{ A}{x_\text{ms}(T)x_\text{ms}(\tau)} \int_{x_0}^{x_0+x_\text{ms}(T)} \exp\left(-\frac{x^2}{2\sigma^2}\right)\left[\exp\left(-\frac{2xx_\text{ms}(\tau) + x_\text{ms}(\tau)^2}{2\sigma^2}\right) -1 \right]dx\\
%	&= \frac{ A}{x_\text{ms}(T)x_\text{ms}(\tau)} \int_{x_0}^{x_0+x_\text{ms}(T)} \exp\left(-\frac{x^2}{2\sigma^2}\right)\left[\exp\left(-\frac{2xx_\text{ms}(\tau) + x_\text{ms}(\tau)^2}{2\sigma^2}\right) -1 \right]dx
%\end{align}

\newpage

\section{Gradient-sensing is the optimal strategy in linear stimulant landscapes}

In linear stimulant landscapes the local gradient already contains complete information about the whole landscape. Therefore, the intuition is that in this case, the solution of the optimal foraging model will be invariant under $ \tau $, i.e. $ \rho_\text{ON}(x,t; \tau) = \rho_\text{ON}(x,t)$. 

To formalise this, we solve the ON model (Eq. (11) in the paper) for $ S_1 = \alpha x $. As derived in Appendix B (Eqs. (B3a) and (B3b) in the paper), the optimality conditions of (11) read 
    \begin{align}
      &  \partial_s m +\nabla\cdot{(m u)}=0,\,\, m(\cdot,0)=\rho(\cdot,k\tau)\tag{B3a}\\
       & \partial_s\phi + u\cdot\nabla\phi + \frac{|u|^2}{2}=0,\,\, \phi(\cdot,\tau)=-\frac{\delta\mathcal{F}[m]}{\delta m}\biggr|_{s=\tau},\tag{B3b}
    \end{align}
with $ u =-\nabla \phi $ and $\mathcal{F}[m] = \int_\Omega \left(m\log m - \text{Pe}_1 m S_1 - \frac{\text{Pe}_2}{2} m(\log|m|\ast m) \right)dx$. Note that since we are interested in the effect of the landscape $S_1$ on the drift velocity, we let $ \text{Pe}_2=0 $ (i.e. the searchers are non-interacting) and $\int_\Omega \rho\log\rho dx = 0$. Considering the HJ equation (B3b) we can then write
\begin{equation}\label{HJB}
	\partial_s\phi - \frac{1}{2}|\nabla\phi|^2=0, \quad \quad \phi(x,\tau)=-\text{Pe}_1 S_1(x) ,
\end{equation}
%
%where we introduced a small parameter $ \epsilon $. This suggests that to find a solution to \eqref{HJB}, we take a perturbative approach and pose the ansatz $ \phi=\phi^0+\epsilon\phi^1+\mathcal{O}(\epsilon^2) $. Expanding in orders of $ \epsilon $, we obtain:
%%
%\begin{subequations}
%\begin{alignat}{2}
%	&\epsilon^0:\quad\partial_s\phi^0 -\frac{1}{2}|\partial_x\phi^0|^2=0, \quad &&\phi^0(\tau,x)=-\text{Pe}_1 S_1(x) \label{eps0}\\
%	&\epsilon^1:\quad\partial_s\phi^1  + \phi^0\partial_x\phi^1 - \phi^1\partial_x \phi^0=0, \quad &&\phi^1(x,\tau)= \log\rho\label{eps1}.
%\end{alignat}
%\end{subequations}
%
Equation~\eqref{HJB} can be solved by the general method of characteristics (\cite{evans}, ch. 3), as follows. Define the parametrisation of space ($x$) and time ($s$) in terms of $ \mu $ and $ \nu $, respectively, and let 
\begin{align*}
	p(\mu,\nu) & :=\partial_s\phi(x(\mu,\nu),s(\mu,\nu)) \\
	q(\mu,\nu) & :=\nabla\phi(x(\mu,\nu),s(\mu,\nu)) \\ 
	z(\mu,\nu) & :=\phi(x(\mu,\nu),s(\mu,\nu)).
\end{align*}
Then we may write \eqref{HJB} as: 
\begin{equation}\label{HJB2}
	\begin{cases}	
		F(x,s,z,p,q)=p-\frac{|q|^2}{2}=0\\
		z(x,\tau) = -\text{Pe}_1 S_1(x)
	\end{cases}
\end{equation}
The characteristic equations and the corresponding boundary conditions are:
\begin{subequations}\label{chareq}
\begin{alignat}{2}
	&\frac{dx}{d\nu}=\frac{\partial F}{\partial q}=-q;\quad &&x(\mu,0)=\mu\\
	&\frac{ds}{d\nu}=\frac{\partial F}{\partial p}=1;\quad &&s(\mu,0)=\tau\\	
	&\frac{dz}{d\nu}= p\frac{\partial F}{\partial p} + q\frac{\partial F}{\partial q}=p-q^2;\quad &&z(\mu,0)=-\text{Pe}_1 S_1(\mu)  \\
	&\frac{dp}{d\nu}=-\frac{\partial F}{\partial s}-p\frac{\partial F}{\partial z}=0;\quad &&p(\mu,0)=\psi_1(\mu)\\
	&\frac{dq}{d\nu}=-\frac{\partial F}{\partial x}-q\frac{\partial F}{\partial z}=0;\quad &&q(\mu,0)=\psi_2(\mu).
\end{alignat}
\end{subequations}
Here $ \psi_1 $ and $ \psi_2 $ are boundary conditions for $ p $ and $ q $, respectively, which must therefore satisfy \eqref{HJB2}
\begin{equation*}
	F(x,s,z,\psi_1,\psi_2)=0,
\end{equation*}
along with the the compatibility condition 
\begin{equation*}
	\frac{d}{d\mu} z(x(\mu,\nu),s(\mu,\nu))\biggr|_{\nu=0} =  \nabla z\frac{\partial}{\partial \mu}x(\mu,\nu)\biggr|_{\nu=0} + \frac{\partial z}{\partial s}\frac{\partial}{\partial \mu}s(\mu,\nu)\biggr|_{\nu=0}.
\end{equation*}
Hence the following two conditions must hold
\begin{equation*}
\begin{cases}
	\frac{d}{d\mu}z(\mu,0)=q \frac{\partial }{\partial\mu}x(\mu,0) + p\frac{\partial }{\partial \nu}s(\mu,0) \\
	\psi_1 - \frac{|\psi_2|^2}{2}=0.
\end{cases}
\end{equation*}
Using the boundary conditions in \eqref{chareq}, we obtain that $ \psi_1=| \partial_\mu S_1(\mu)|^2/2 $ and $\psi_2 = - \partial_\mu S_1(\mu)$. Then, integrating the characteristic equations \eqref{chareq} we obtain
\begin{align}
	x&= \partial_\mu S_1(\mu)\nu+\mu\\
	s&=\tau-\nu\\
	z&=\frac{| \partial_\mu S_1(\mu)|^2}{2}\nu-\text{Pe}_1S_1(\mu),
\end{align}
which need to be inverted to express $ \mu,\nu,z $ in terms of $ x,s $. 

\paragraph*{Particularisation for linear stimulant landscapes ---}

For linear landscapes $ S_1(\mu)= \alpha \mu $, it is possible to invert the above equations to obtain $ \phi=z(x,s)=-\alpha\text{Pe}_1 x + f(s)$. %Since $ S_0(x) $ is a linear function, we have $ \phi^n=0 $ for $n >0$. 
Using the optimality condition, we find that the mean chemotactic velocity is $ u(x,s) = -\nabla\phi = \alpha\text{Pe}_1 $, independent of $ \tau $, as we set out to show.

In conclusion, under linear stimulant landscape local gradient information characterises the field globally and there is no benefit from having a finite spatial horizon.

\newpage

\section{Supplementary figures}

\begin{figure}[h!]
\centering
\includegraphics[width=0.7\textwidth]{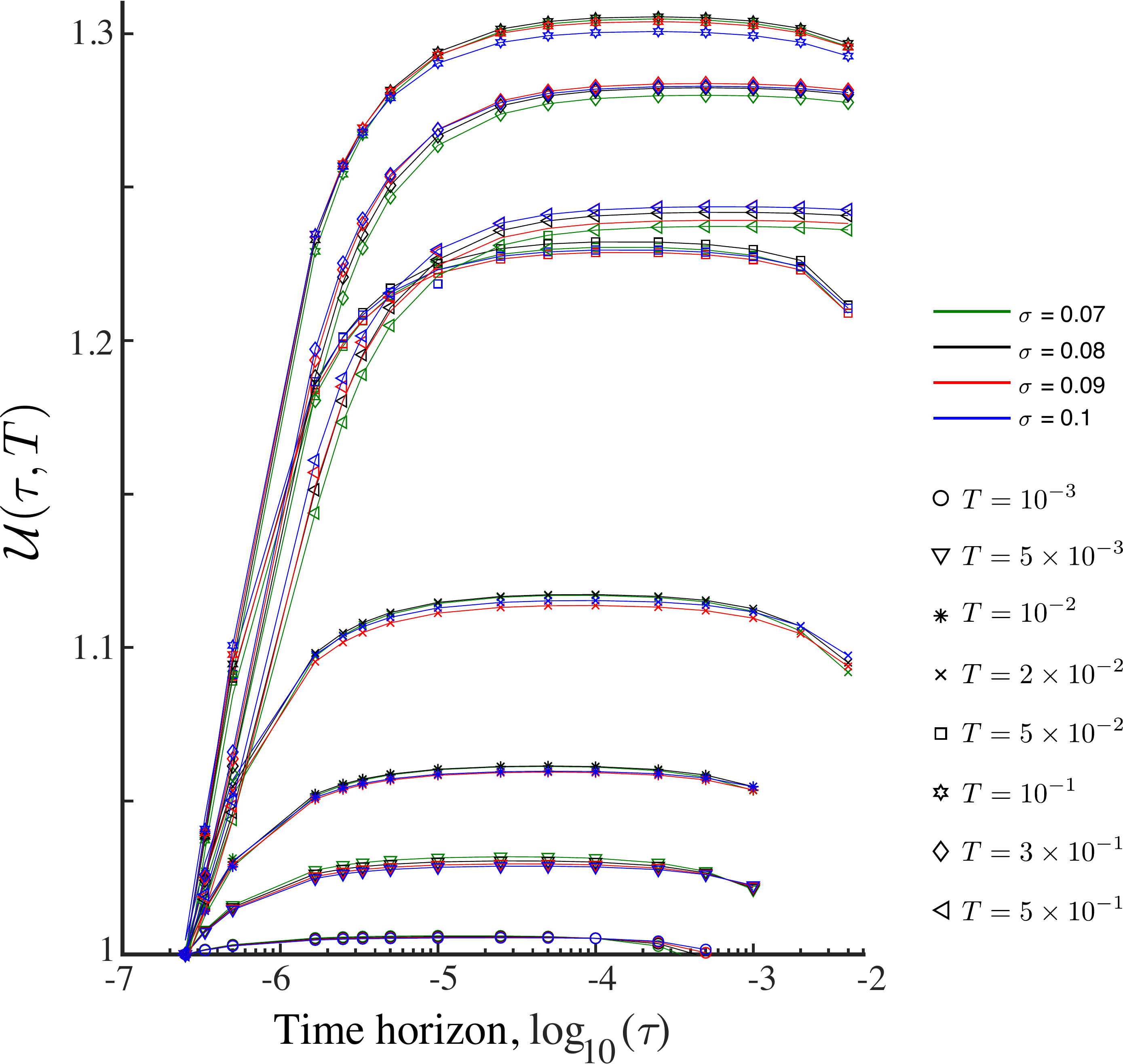}
\caption{Relative population fitness against time horizon at fixed search times $ T $ and different environmental length scales $ \sigma $ (see legend). Markers are obtained by renormalising the P\'{e}clet number as in (23), simulating a population evolution in the given stimulant landscape using the ON model (Eq. (11) in the paper) and computing $ \mathcal{U}(\tau,T) $ from (17). The maxima of the curves $ \mathcal{U}^*(T) $ are used to compute Fig. 4d in the paper. }
\end{figure}

\bibliography{refs}
 
\end{document}